\documentclass{article}
\usepackage[utf8]{inputenc}

\usepackage{tikz}
\definecolor{colorSFPLAIC}{HTML}{cc0000}
\definecolor{colorSFPLBIC}{HTML}{f1c232}
\definecolor{colorPL}{HTML}{6aa84f}
\definecolor{colorPPL}{HTML}{3d85c6}

\usepackage[unicode,linktoc=all,hidelinks]{hyperref}
\usepackage[a4paper, margin=2cm]{geometry}
\usepackage{mathtools}

\usepackage{amsmath}
\usepackage{amssymb}
\usepackage{amsthm}
\usepackage{bm}
\usepackage{bbm}

\DeclareSymbolFont{cmletters}{OT1}{cmr}{m}{n}
\DeclareMathSymbol{\Ups}{\mathalpha}{cmletters}{"7}

\usepackage{verbatim}

\usepackage{diagbox}
\usepackage{float} 
\usepackage{adjustbox}
\usepackage{threeparttable}
\usepackage{booktabs}

\usepackage{algorithm}
\usepackage{algorithmic}
\usepackage[english]{babel}
\usepackage{natbib}

\usepackage{graphicx}

\usepackage[title]{appendix}

\newtheorem{theorem}{Theorem}

\usepackage{algorithm}
\usepackage{algorithmic}
\title{A Statistical Interpretation of Multi-Item Rating and Recommendation Problems}
\author{Sjoerd Hermes$^{1,2}$}
\date{%
    $^1$ Mathematical and Statistical Methods, Wageningen University\\%
    $^2$ Plant Production Systems, Wageningen University\\
}

\begin{document}

\maketitle
\begin{abstract}
\noindent Ordinal user-provided ratings across multiple items are frequently encountered in both scientific and commercial applications. Whilst recommender systems are known to do well on these type of data from a predictive point of view, their typical reliance on large sample sizes and frequent lack of interpretability and uncertainty quantification limits their applicability in inferential problems. Taking a fully Bayesian approach, this article introduces a novel statistical method that is designed with interpretability and uncertainty quantification in mind. Whilst parametric assumptions ensure that the method is applicable to data with modest sample sizes, the model is simultaneously designed to remain flexible in order to handle a wide variety of situations. Model performance, i.e.\ parameter estimation and prediction, is shown by means of a simulation study, both on simulated data and against commonly used recommender systems on real data. These simulations indicate that the proposed method performs competitively. Finally, to illustrate the applicability of the proposed method on real life problems that are of interest to economists, the method is applied on speed dating data, where novel insights into the partner preference problem are obtained. An R package containing the proposed methodology can be found on \url{https://CRAN.R-project.org/package=StatRec}.
\end{abstract}

\noindent%

\section{Introduction}

Rating data frequently occur not only within a variety of scientific disciplines such as computer science (Koren et al., \citeyear{koren2009matrix}), economics (Fisman et al., \citeyear{fisman2008racial}), marketing (Fainmesser et al., \citeyear{fainmesser2021ratings}) and psychology (Busing et al., \citeyear{busing2005avoiding}), but also in non-scientific  applications such as e-commerce (Danescu-Niculescu-Mizil et al., \citeyear{danescu2009opinions}), media streaming (Bell et al., \citeyear{bell2007lessons}) and travel (Mariani \& Borghi, \citeyear{mariani2018effects}). Whilst, in their most general format, these data consist of observations made by users that describe some quality of the item that is being rated, this paper focusses on the subset of rating data where each rating takes some ordinal (ordered categorical) value in $\{1,2,\ldots,k\} = \mathcal{K}$. %Commonly, these data are studied by accompanying the ratings with a set of covariates that express some properties of the observations, with the aim of uncovering the dependency between the ratings and the covariates in order to make predictions or acquire understanding of the rating process (Fisman et al., \citeyear{fisman2008racial}; Jones et al., \citeyear{jones2015empirical}). 
Commonly, a set of covariates that express some properties of the observations is included in the analysis of the rating data, in order to acquire understanding of the rating process or make predictions (Fisman et al., \citeyear{fisman2008racial}; Jones et al., \citeyear{jones2015empirical}). 

To study this dependency between ordinal ratings and covariates, a variety of statistical models have been proposed, where the class of ordinal regression models (McCullagh, \citeyear{mccullagh1980regression}; Agresti \citeyear{agresti2012categorical}) has been the most popular. These models have been extended in various ways, such as to allow for multivariate ordinal data (Lawrence et al., \citeyear{lawrence2008bayesian}), sparsity (Chen et al., \citeyear{chen2023bayesian}), quantile regression (Rahman, \citeyear{rahman2016bayesian}) and subgroups of observations (DeYoreo et al., \citeyear{deyoreo2017bayesian}). However, to the best of my knowledge, all statistical methods developed for the analysis of rating data assume that the observed ratings concern the same item. In many scientific and non-scientific settings, ratings are observed across a variety of items. Consider, for example, an electronic merchant that sells books. After purchasing a book (the item), a consumer (the user) can provide a rating for that book, which the merchant can collect, relate to various covariates describing the consumer and finally use these to entice the consumer into purchasing other books. However, given a set of covariates describing some consumer, the predicted ratings for different books will be the same when using any of the existing statistical methods for ordinal data, as these have no way of differentiating between the different books. This is unrealistic, as any observed rating should not only depend on the individual that expressed that rating, but also on the item that is being rated (Rahimi et al., \citeyear{rahimi2022know}). Whilst a potential workaround to this complication consists of using a multivariate ordinal model (Lawrence et al., \citeyear{lawrence2008bayesian}; Bao \& Hanson, \citeyear{bao2015bayesian}; DeYoreo \& Kottas, \citeyear{deyoreo2018bayesian}), this class of models is unsuitable for two reasons: they assume that each individual rates each item, which is costly in scientific experiments, and unrealistic in non-scientific applications, and they postulate a separate coefficient vector for the effect of the covariates pertaining to the user on each item, but do not use any information inherent to the items that are rated. 

Recommender systems do model the interaction between users and items. They comprise a collection of machine learning techniques that aim to provide accurate and personalized recommendations of items to users  (LeBlanc et al., \citeyear{leblanc2024recommender}). The similarity with the abovementioned scenario is especially striking for the subset of recommender systems that utilize a function of both item and user covariates to express an observed rating (Chu \& Park, \citeyear{chu2009personalized}; Park \& Chu, \citeyear{park2009pairwise}; Agarwal et al., \citeyear{agarwal2010fast}). This similarity notwithstanding, these methods are not developed with ordinal-type rating data in mind, but instead assume either binary or continuous rating data. Moreover, they might be overly restrictive as the assumption is made that the observed rating is a product of interactions between the item and user covariates, rather than allowing for more parsimonious structures, such as additive ones. Alternatively, the class of latent factor recommender systems (Dai et al., \citeyear{dai2021scalable}; LeBlanc et al., \citeyear{leblanc2024recommender}) do provide a flexible manner of modelling the ratings: these models assume the existence of latent user and item factors, but do not take any explicit covariate information into account, hindering interpretability. One exception to the lack of interpretability of latent factors models is the work of Zhu et al., (\citeyear{zhu2016personalized}), which incorporates both coefficients corresponding to user and item variables, as well as latent factors. Nevertheless, in many applications, both scientific and non-scientific, having the ability to quantify the uncertainty surrounding the parameter estimates is not just a philosophical matter or a theoretical nicety, but an essential element of obtaining understanding of a given phenomenon or in decision making processes (Bhattacharya et al., \citeyear{bhattacharya2015dirichlet}; Cai et al., \citeyear{cai2023statistical}). Recommender systems typically lack this important feature. Whilst research into uncertainty quantification for recommender systems has been gaining some traction recently (Wang et al., \citeyear{wang2018confidence}; Coscrato \& Bridge, \citeyear{coscrato2023estimating}; Wang \& Joachims, \citeyear{wang2023uncertainty}; Ma et al., \citeyear{ma2024statistical}), these novel works still have the same drawbacks as the other recommender systems do mentioned above. Finally, recommender systems typically require large amounts of data (Sun \& Peng, \citeyear{sun2022survey}), due to their nonparametric nature. Whilst commercial applications might consist of many observations, scientific applications do not, due to budgetary or time constraints. As such, there is ample added value in a statistical method that allows practitioners to conduct inference on multi-item rating and recommendation problems without relying on big data. 
% Note that the work of ma2024statistical, is quite good, but out method is fully Bayesian, and therefore allows for credibility intervals on the coefficients, latent factors and predicted ratings. Moreover, it is specifically tailored for ordinal rating data, it can handle item covariates and it can handle a large amount of covariates due to the shrinkage.

Therefore, this paper combines the best aspects of both the ordinal regression models and recommender systems, resulting in an easy to interpret and flexible statistical method than can model rating data across multiple items, whilst simultaneousely providing uncertainty quantifications surrounding the parameter estimates and predicted values and allowing for inference using small to moderate sample sizes by relying on a set of parametric assumptions. Even though a Bayesian approach to statistical inference is the most natural way to quantify uncertainty, when the data generating mechanism is postulated to resemble an ordinal regression method, sampling from the posterior distribution is highly nontrivial (Albert \& Chib, \citeyear{albert1993bayesian}). Nevertheless, by manipulating the term within the link function found in the likelihood, the Pólya-Gamma framework (Polson et al., \citeyear{polson2013bayesian}) can be utilized, thereby enabling efficient and straightforward sampling from the posterior distribution. Moreover, depending on the assumed complexity of the underlying phenomenon, the proposed method allows for both linear and bilinear terms in the link function, increasing the flexibility of the method. To further improve the predictive capacity and interpretability of the proposed method, it allows for both latent factors and sparsity inducing priors on the coefficients corresponding to the item and user covariates.

This paper proceeds in Section \ref{A multi-item rating statistical model} by introducing the underlying statistical method. Subsequently, Section \ref{Bayesian inference} covers the Bayesian estimation framework required to estimate the parameters of interest. In Section \ref{Simulation study} a simulation study is presented to show how well the method performs on both model-based simulations, as well as real-world data. This section is followed by Section \ref{Real data application}, which provides an application on data pertaining to a speed dating experiment. The paper concludes in Section \ref{Conclusion and discussion} with a conclusion and discussion for future research directions. 

\section{A multi-item rating statistical model} \label{A multi-item rating statistical model}

Let rating data $\bm{R} = (r_{ij})_{n \times m}$ be given. These data consists of $n$ users and $m$ items, such that each user $i$ provides a rating on items $\mathcal{M}_i$, $\mathcal{M}_i \subsetneq \{1,2,\ldots,m\}$ ,where $r_{ij} \in \mathcal{K} \Leftrightarrow j \in \mathcal{M}_i$ and $r_{ij}  = 0$ otherwise. From this notation, it becomes apparent that $|\mathcal{M}_i| < m$ for at least one $i$, or stated otherwise, that not all users provide a rating for all items. This situation is frequently encountered 
in real-world data (LeBlanc et al., \citeyear{leblanc2024recommender}), given that with an increasing number of items available to users, it becomes less likely that each user is able to provide a rating for all items. Nevertheless, the proposed method is also applicable to data where $|\mathcal{M}_{i}| = m$ for some or all $i$.
\\
\\
\noindent Observed ratings are expected to depend on properties relating to the user that rates an item, as well as properties relating to that item, such that $r_{ij} = f(\bm{x}_{i}, \bm{y}_{j})$, where $\bm{x}_i$ and $\bm{y}_j$ are $p$- and $q$-dimensional vectors of covariates related to the $i$th user and $j$th item respectively, such that $\bm{X} \in \mathbb{R}^{n \times p}$ and $\bm{Y} \in \mathbb{R}^{m \times q}$. This type of relationship has been proposed before in the context of recommender systems (Chu \& Park, \citeyear{chu2009personalized}; Park \& Chu, \citeyear{park2009pairwise}; Agarwal et al., \citeyear{agarwal2010fast}; Zhu et al. \citeyear{zhu2016personalized}). 

As a first step towards a statistical model for rating data, two variants of the predictor are proposed:
\begin{equation}
\label{eq:eta_def}
\eta_{ij}^{(l)} = \begin{bmatrix}
\bm{x}_{i} & \bm{y}_{j}
\end{bmatrix}^T\bm{b}, \quad \text{ and } \quad \eta_{ij}^{(b)} = \bm{x}_{i}^T\bm{B}\bm{y}_{j},
\end{equation}
with $\bm{b} \in \mathbb{R}^{p + q}$ for $\eta_{ij}^{(l)}$ and $\bm{B} \in \mathbb{R}^{p \times q}$ for $\eta_{ij}^{(b)}$. As such, $\eta_{ij}^{(l)}$ represents a linear predictor, where the effects of the user and item covariates are additive. Conversely, $\eta_{ij}^{(b)}$ represents a bilinear predictor where the user and item covariates interact by means of their weighted product. To elaborate on this, for any $\bm{B} = \left(b_{p'q'}\right)_{p \times q}$, the value of $b_{p'q'}$ is interpreted as the weight assigned to the interaction between the $p'$th and $q'$th variables in $\bm{X}$ and $\bm{Y}$ respectively. Whilst this results in a heavily parameterised model, the ability of capturing such user-item interactions is both of interest to researchers developing recommender systems (Chu \& Park, \citeyear{chu2009personalized}; Park \& Chu, \citeyear{park2009pairwise}; Agarwal et al., \citeyear{agarwal2010fast}), as well as to other scientific fields such as economics, where the much studied problem of partner preferences also depends on the interaction of many characteristics reflecting two potential partners (Sprecher et al., \citeyear{sprecher1991effect}; Fisman et al., \citeyear{fisman2008racial}; Hitsch et al., \citeyear{hitsch2010matching}; \citeyear{hitsch2010makes}). Section \ref{Real data application}, shows that this partner preference problem can be framed by means of a more typical multi-item rating framework. The choice for either functional form depends on both the assumed complexity of the problem as well interests in specific effects: if the practitioner supposes that the observed ratings arise from a complex interplay between the covariates pertaining to both items and users, the bilinear method is appropriate. If not, the linear alternative should be applied. 
\\
\\
For the remainder of this article, the methodology is presented using the bilinear notation. Nevertheless, all results are generalisable to the linear predictor provided in Equation (\ref{eq:eta_def}). Let
\begin{equation}
\label{eq:g_def}
g\left(\eta_{ij}^{(b)}\right) = \frac{1}{1 + \exp\left(-\bm{x}_{i}^T\bm{B}\bm{y}_{j}\right)},
\end{equation}
such that $g: \mathbb{R} \rightarrow [0,1]$ is the standard logistic function. Then, even though $g$ provides a mapping to $[0,1]$ instead of a mapping to $\mathcal{K}$, a shifted binomial distribution is applied to both ensure that $r_{ij} \in \mathcal{K}$ for observed $r_{ij}$, as well as to turn this into a probabilistic rather than a deterministic framework. The shifted binomial distribution, with a shift of 1, only has positive support for $r_{ij} \in \mathcal{K}$. Consequently, the probability of observing $r_{ij}$ is given by
\begin{equation}
\begin{gathered}
\label{eq:probrij}
\mathbb{P}(r_{ij}|\bm{B}) = \binom{k - 1}{r_{ij} - 1}p_{ij}^{r_{ij} - 1}\left(1 - p_{ij}\right)^{k - r_{ij}}\\
= \binom{k - 1}{r_{ij} - 1}\frac{\exp\left(\bm{x}_{i}^T\bm{B}\bm{y}_{j}\right)^{r_{ij} - 1}}{\left[1 + \exp\left(\bm{x}_{i}^T\bm{B}\bm{y}_{j}\right)\right]^{k - 1}},
\end{gathered}
\end{equation}
where the equality $g\left(\eta_{ij}^{(b)}\right) = p_{ij}$ is used. By replacing $k - 1$ with $k$ and $r_{ij} - 1$ with $r_{ij}$ in Equation (\ref{eq:probrij}), the proposed method is also suitable for ratings taking values in $\{0\}\cup \mathcal{K}$. Additionally, by replacing $k$ with $k_{ij}$, the method can incorporate differing maximum ratings for each user and item. Finally, by replacing the shifted binomial distribution with a Bernoulli distribution, the method can accommodate binary data, such as purchasing data (bought, not bought), preference data (like, dislike) or binary rating data, without complicating the parameter estimation, as a consequence of the flexibility of the proposed Bayesian inferential procedure (see Section \ref{Bayesian inference}). 

\subsection{Extending the model} \label{Latent factors}
In order to estimate effects that cannot be captured by the covariates, such as individual user and item biases, the linear predictors in Equation (\ref{eq:probrij}) can be extended with latent factors such that
\begin{equation}
\label{eq:eta_def_lat}
\eta_{ij}^{(l)} = \begin{bmatrix}
\bm{x}_{i} & \bm{y}_{j}
\end{bmatrix}^T\bm{b} + \bm{u}_{i}\bm{v}_{j}^T, \quad \text{ and } \quad \eta_{ij}^{(b)} = \bm{x}_{i}^T\bm{B}\bm{y}_{j} + \bm{u}_{i}\bm{v}_{j}^T,
\end{equation}
with latent user factors $\bm{U} \in \mathbb{R}^{n \times l}$ and latent item factors $\bm{V} \in \mathbb{R}^{m \times l}$, such that $\bm{F} = \bm{U}\bm{V}^T$ is a low rank matrix. Given that Equation (\ref{eq:eta_def_lat}) requires either $pq + l(n+m)$ or $p+q + l(n+m)$ parameters to be estimated, depending on whether a linear or bilinear model is assumed, $\bm{U}$ and $\bm{V}$ are assumed to be sparse. Despite this assumption, additional assumptions or constraints are required to identify $\bm{U}$ and $\bm{V}$ for $l>1$. However, latent factor models encountered in the recommender system literature typically make no attempt at identifying these matrices (Agarwal et al., \citeyear{agarwal2010fast}; Agarwal \& Chen, \citeyear{agarwal2016statistical}; Liu et al., \citeyear{liu2017collaborative}; Fang et al., \citeyear{fang2021probabilistic}; Ma et al., \citeyear{ma2024statistical}), as $\bm{F}$ is identifiable, which is sufficient for their aim: predicting the missing values in $\bm{R}$. In the proposed method, no identifiability constraints are imposed on $\bm{U}$ and $\bm{V}$, as these have no real interpretation for $l > 1$. Whilst $l$ is fixed and assumed known for the remainder of this article, in practice, this quantity can be determined by, for example, an information criterion (Hirose et al., \citeyear{hirose2011bayesian}; Bai \& Wang, \citeyear{bai2015identification}) or cross-validation approaches (Lu et al., \citeyear{lu2017comparison}). Finally, predicting the ratings for new users and items, commonly referred to as the “cold start problem” (LeBlanc et al., \citeyear{leblanc2024recommender}), is nontrivial for latent variable models, as the estimated latent variables cannot be easily generalized to new users and items. However, given that both $\bm{U}$ and $\bm{V}$, and as a (probable) consequence $\bm{F}$, are assumed to be sparse, the probability that the rating is only a function of the user and item covariates is substantial. Therefore, ratings for new users and items can be predicted using the predictors in Equation (\ref{eq:eta_def}). 
\\
\\
In addition to imposing sparsity on $\bm{U}$ and $\bm{V}$, sparsity can also be imposed on $\bm{B}$ (c.f.\ Agarwal \& Chen, \citeyear{agarwal2016statistical}). Besides improvements related to the model interpretability, sparsity can also improve predictive performance whenever the data consists of relatively many (irrelevant) user and item covariates (Carvalho et al., \citeyear{carvalho2009handling}). To allow for the application of the proposed method across a wide range of situations, sparsity inducing priors on $\bm{B}$ can be used instead of the regular non-sparsity inducing ones, as will be shown in Section \ref{Bayesian inference}. 

\section{Bayesian inference} \label{Bayesian inference}
The model parameters are estimated using a Bayesian approach that is free of any hyperparameter specification. First, the parameter estimation for the model without latent variables, see Equation (\ref{eq:eta_def}), is illustrated, followed by an illustration of the estimation process of the model with latent variables, see Equation (\ref{eq:eta_def_lat}).
\\
\\
The first step of any Bayesian inferential procedure consists of formulating the likelihood. Assuming that the $r_{ij}$ are i.i.d., the following likelihood is obtained
\begin{equation}
\begin{gathered}
\label{eq:lik}
\mathbb{P}\left(\bm{R}|\bm{B}\right) = \prod_{i = 1}^n\prod_{j \in \mathcal{M}_i}\mathbb{P}\left(r_{ij}|\bm{B}\right)\\
%\left(\bm{B}|\bm{R}, \bm{X}, \bm{Y}\right) = \prod_{i = 1}^n\prod_{j \in \mathcal{M}_i}\mathbb{P}\left(r_{ij}|\bm{B}\right)\\
= \prod_{i = 1}^n\prod_{j \in \mathcal{M}_i}\binom{k - 1}{r_{ij} - 1}\frac{\exp\left(\bm{x}_{i}^T\bm{B}\bm{y}_{j}\right)^{r_{ij} - 1}}{\left[1 + \exp\left(\bm{x}_{i}^T\bm{B}\bm{y}_{j}\right)\right]^{k - 1}}.
\end{gathered}
\end{equation}
By Bayes theorem, the posterior  $\mathbb{P}(\bm{B}|\bm{R})$ is given by $\mathbb{P}(\bm{B}|\bm{R}) \propto \mathbb{P}\left(\bm{R}|\bm{B}\right)\pi(\bm{B})$, where $\pi(\bm{B})$ is a prior on $\bm{B}$. Unfortunately, sampling from this posterior is highly nontrivial, due to the inconvenient functional form of the likelihood. As the Metropolis-Hastings algorithm (Metropolis et al., \citeyear{metropolis1953equation}; Hastings, \citeyear{hastings1970monte}), tends to be computationally inefficient for high-dimensional parameter estimation, various authors have tried to speed up Bayesian inference for the logistic regression model using data augmentation techniques (Groenewald \& Lucky, \citeyear{groenewald2005bayesian}; Holmes \& Held, \citeyear{held2006bayesian}; Frühwirth-Schnatter \& Frühwirth, \citeyear{fruehwirth2007auxiliary}). Noting that these methods are typically either approximate or relatively complex, Polson et al.\ (\citeyear{polson2013bayesian}) also developed a data augmentation-based framework. Their approach assumes a binomial-type likelihood, and by augmenting the data with Pólya-Gamma random variables, their method allows for efficient Bayesian inference for the logistic regression model. Given the method's efficiency and flexibility, it seen many successful applications of various types of logistic models, such as hierarchical negative binomial models (Neelon et al., \citeyear{neelon2019bayesian}) Poisson models (D'Angelo \& Canale, \citeyear{d2023efficient}) and extensions to imbalanced categorical data (Zens et al., \citeyear{zens2023ultimate}). The Pólya-Gamma data augmentation scheme is also used in this paper.
\\
\\
As an aside, it should be noted that the likelihood presented in Equation (\ref{eq:lik}) is not identifiable without additional restrictions. Whilst informative priors can circumvent this problem, the following Theorem is presented to ensure that the model is not used under the wrong circumstances.

\begin{theorem}
The likelihood presented in Equation (\ref{eq:lik}) is identifiable if $\text{rank}\left(\bm{X}\right) = p$ and $\text{rank}\left(\bm{Y}\right) = q$.
\end{theorem}

\noindent A proof of this result is provided in the supplementary material.

\subsection{Estimation using Pólya-Gamma data augmentation} \label{Pólya-Gamma data augmentation} 
In this section, the key concepts of the Pólya-Gamma data augmentation are briefly explained, before the posterior sampling scheme of the proposed method is introduced. Central to the data augmentation method proposed by Polson et al.\ (\citeyear{polson2013bayesian}) is the Pólya-Gamma random variable $\omega$. Formally stated: a random variable $\omega$ follows a Pólya-Gamma distribution with parameters $a > 0$ and $b = 0$, if
\begin{equation}
%\omega \sim \text{PG}(a,b) \stackrel{D}{=} \frac{1}{2\pi^2}\sum_{s=1}^\infty \frac{\text{Gamma}(a,1)}{(s - 1/2)^2 + b^2/(4\pi^2)},
\omega \sim \text{PG}(a,0) \stackrel{D}{=} \frac{1}{2\pi^2}\sum_{s=1}^\infty \frac{\text{Gamma}(a,1)}{(s - 1/2)^2},
\end{equation}
for mutually independent $\text{Gamma}(a,1)$ random variables. The main result of Polson et al.\ (\citeyear{polson2013bayesian}) is that binomial-type likelihoods parameterized by a logistic function can be represented as mixtures of Gaussians with respect to a Pólya-Gamma distribution. Modifying the identity of Polson et al.\ (\citeyear{polson2013bayesian}), then, for any $\omega_{ij} \sim \text{PG}(k-1,0)$, whose density is denoted by $\mathbb{P}\left(\omega_{ij}|k-1,0\right)$ the following equality
\begin{equation}
\label{eq:psidentity}
\frac{\exp\left[\bm{z}_{ij}^T\text{vec}(\bm{B})\right]^{r_{ij} - 1}}{\left\{1 + \exp\left[\bm{z}_{ij}^T\text{vec}(\bm{B})\right]\right\}^{k - 1}} = \frac{1}{2^{k-1}}\exp\left[\kappa_{ij} \bm{z}_{ij}^T\text{vec}(\bm{B})\right]\int_{0}^\infty \exp\left\{-\frac{1}{2}\omega_{ij}\left[\bm{z}_{ij}^T\text{vec}(\bm{B})\right]^2\right\}\mathbb{P}\left(\omega_{ij}|k-1,0\right)d\omega_{ij}
\end{equation}
holds, where $\kappa_{ij} = r_{ij} - (k + 1)/2$, and, to simplify notation, $\bm{z}_{ij} = \bm{y}_{j} \otimes \bm{x}_{i}$ is used, with $\bm{x}_{i}^T\bm{B}\bm{y}_{j} = \bm{z}_{ij}^T\text{vec}(\bm{B})$. The same equality as in Equation (\ref{eq:psidentity}) holds for the linear model, where instead $\bm{z}_{ij} = \begin{bmatrix}
\bm{x}_{i} & \bm{y}_{j}
\end{bmatrix}^T$ and $\text{vec}(\bm{B}) = \bm{b}$. Finally, the following conditional distribution
\begin{equation}
\mathbb{P}\left(\omega_{ij}|k-1, \bm{B}\right) = \frac{\exp\{-\frac{1}{2}\omega_{ij}[\bm{z}_{ij}^T\text{vec}(\bm{B})]^2\}\mathbb{P}(\omega_{ij}|k-1, 0)}{\int_{0}^\infty\exp\{-\frac{1}{2}\omega_{ij}[\bm{z}_{ij}^T\text{vec}(\bm{B})]^2\}\mathbb{P}(\omega_{ij}|k-1, 0)d\omega_{ij}},
\end{equation}
arises from an “exponential tilting” of the $\text{PG}(k-1,0)$ density. Therefore, the conditional posterior for the $\omega_{ij}$ is given by
\begin{equation}
\omega_{ij}|\bm{B} \sim \text{PG}\left[k-1, \bm{z}_{ij}^T\text{vec}\left(\bm{B}\right)\right].
\end{equation}
On the other hand, it can be seen that the full conditional distribution of $\text{vec}\left(\bm{B}\right)$ is given by
\begin{equation}
\begin{gathered}
\mathbb{P}\left[\text{vec}\left(\bm{B}\right)|\bm{\Omega}\right] \propto \pi\left[\text{vec}\left(\bm{B}\right)\right]\prod_{i = 1}^n\prod_{j \in \mathcal{M}_i}\mathbb{P}\left(r_{ij}|\bm{B}\right)\mathbb{P}\left(\omega_{ij}|r_{ij},\bm{B}\right)\\
%= \pi\left[\text{vec}\left(\bm{B}\right)\right]\prod_{i = 1}^n\prod_{j \in \mathcal{M}_i}\frac{\exp\left[\bm{z}_{ij}^T\text{vec}\left(\bm{B}\right)\right]^{r_{ij} - 1}}{\left\{1 + \exp\left[\bm{z}_{ij}^T\text{vec}\left(\bm{B}\right)\right]\right\}^{k - 1}}\mathbb{P}\left(\omega_{ij}|r_{ij},\bm{B}\right)\\
= \pi\left[\text{vec}\left(\bm{B}\right)\right]\exp\left\{-\frac{1}{2}\left[\text{vec}\left(\bm{\Xi}\right) - \bm{Z}\text{vec}\left(\bm{B}\right)\right]^T\bm{\Omega}\left[\text{vec}\left(\bm{\Xi}\right) - \bm{Z}\text{vec}\left(\bm{B}\right)\right]\right\},
\end{gathered}
\end{equation}
where $\bm{\Xi} = (\xi_{ij})_{n \times m}, \xi_{ij} = \kappa_{ij}/\omega_{ij}, \bm{\Omega} = \text{diag}(\omega_{11},\omega_{21},\ldots,\omega_{n1},\ldots,\omega_{1m},\omega_{2m},\ldots,\omega_{nm})$, with $\omega_{ij} \neq 0 \Leftrightarrow j \in \mathcal{M}_{i}$ and $\kappa_{ij} \neq 0 \Leftrightarrow j \in \mathcal{M}_{i}$. In addition, for the bilinear model, $\bm{Z} = \bm{Y} \otimes \bm{X}$, whereas for the linear model $\bm{Z} = \left[\sum_{i = 1}^m \bm{e}_i^{(m)} \otimes \bm{X}, \sum_{j = 1}^n \bm{Y} \otimes \bm{e}_j^{(n)} \right]$, with $\bm{e}_i^{(m)}$ and $\bm{e}_j^{(n)}$ being standard basis vectors in $\mathbb{R}^m$ and $\mathbb{R}^n$ respectively. 
\\
\\
\noindent Assuming that $\pi\left[\text{vec}\left(\bm{B}\right)\right] = N_{pq}(\bm{\mu}_0, \bm{\Sigma}_0)$, the resulting conditional posterior of $\text{vec}\left(\bm{B}\right)$ is given by
\begin{equation}
\text{vec}\left(\bm{B}\right)| \bm{R}, \bm{\Omega} \sim N_{pq}(\bm{\mu}_{\text{vec}\left(\bm{B}\right)}, \bm{\Sigma}_{\text{vec}\left(\bm{B}\right)}),
\end{equation}
where
\begin{equation*}
\bm{\Sigma}_{\text{vec}\left(\bm{B}\right)} = \left(\bm{\Sigma}_0^{-1} + \bm{Z}^T\bm{\Omega}\bm{Z}\right)^{-1}\quad \text{ and } \quad \bm{\mu}_{\text{vec}\left(\bm{B}\right)} = \bm{\Sigma}_{\text{vec}\left(\bm{B}\right)}\left[\bm{\Sigma}_0^{-1}\bm{\mu}_0 + \bm{Z}^T\text{vec}\left(\bm{\kappa}\right)\right],
\end{equation*}
where in practice $\bm{\mu}_0$ and $\bm{\Sigma}_0$ are set to the zero vector and identity matrix respectively. For computational efficiency, the entries corresponding to any unit-item combination $i$,$j$ for which $r_{ij} = 0$ are removed from $\bm{Z}$, $\bm{\kappa}$ and $\bm{\Omega}$. The savings in computation time are especially large for sparse $\bm{R}$, with large $n$ and $m$. 
\\
\\
Algorithm \ref{alg:polgam} summarizes the Gibbs sampling procedure for the posterior of interest $\mathbb{P}\left[\text{vec}\left(\bm{B}\right)| \bm{R}, \bm{\Omega}\right]$.
\begin{algorithm}[H]
\caption{Posterior sampling scheme using Pólya-Gamma data augmentation}\label{alg:polgam}
    \hspace*{\algorithmicindent} \textbf{Input:} Data $\bm{R}$, user covariates $\bm{X}$, item covariates $\bm{Y}$, priors $\bm{\mu}_0$ and $\bm{\Sigma}_0$ and number of iterations $T$.\\
    \hspace*{\algorithmicindent} \textbf{Output:} Posterior samples $\left\{\bm{B}^{[1]},\bm{B}^{[2]},\ldots,\bm{B}^{[T]}\right\}$.
  \begin{algorithmic}[1]
        \STATE \textbf{Initialize:} Set $\bm{B}^{[0]} = \bm{0}_{pq}$ and for all $i,j$ set $\kappa_{ij} = r_{ij} - (k+1)/2$ if $j \in \mathcal{M}_{i}$ and $\kappa_{ij} = 0$ otherwise
            \FOR{$t = 1$ to $T$}
      		\FOR{$i = 1$ to $n$}
		\FOR{$j = 1$ to $m$}
		\IF{$j \in \mathcal{M}_{i}$}
		\STATE Draw $\omega_{ij}^{[t]}|\bm{B}^{[t-1]} \sim \text{PG}\left[k-1, \bm{z}_{ij}\text{vec}\left(\bm{B}^{[t-1]}\right)\right]$
		%\STATE Set $\xi_{ij}^{[t]} = \kappa_{ij}/\omega_{ij}^{[t]}$
		\ELSE
		\STATE Set $\omega_{ij}^{[t]} = 0$
		%\STATE Set $\xi_{ij}^{[t]} = 0$
		\ENDIF
		\ENDFOR
		\ENDFOR
	\STATE Set $\bm{\Sigma}_{\text{vec}\left(\bm{B}\right)}^{[t]} = \left(\bm{\Sigma}_0^{-1} + \bm{Z}^T\bm{\Omega}^{[t]}\bm{Z}\right)^{-1}$
	%\STATE Set $\bm{\mu}_{\bm{\beta}}^{[t]} = \bm{\Sigma}_{\bm{\beta}}^{[t]}\left[\bm{\Sigma}_0^{-1}\bm{\mu}_0 + \bm{Z}^T\bm{\Omega}^{[t]}\text{vec}\left(\bm{\Xi}^{[t]}\right)\right]$	
		\STATE Set $\bm{\mu}_{\text{vec}\left(\bm{B}\right)}^{[t]} = \bm{\Sigma}_{\bm{\beta}}^{[t]}\left[\bm{\Sigma}_0^{-1}\bm{\mu}_0 + \bm{Z}^T\text{vec}\left(\bm{\kappa}\right)\right]$	
        \STATE Draw $\text{vec}\left(\bm{B}^{[t]}\right)| \bm{R}, \bm{\Omega}^{[t]} \sim N_{pq}(\bm{\mu}_{\text{vec}\left(\bm{B}\right)}^{[t]}, \bm{\Sigma}_{\text{vec}\left(\bm{B}\right)}^{[t]})$
      \ENDFOR
  \end{algorithmic}
\end{algorithm}

\subsection{Bayesian estimation of the extended model} \label{Bayesian inference with latent factors}
Given that the latent factors extend the linear predictor in Equation (\ref{eq:eta_def}), with some matrix algebra, Pólya-gamma data augmentation can also be used to estimate $\bm{F}$. First, for all $i,j$ Pólya-gamma random variables are sampled as 
\begin{equation}
\omega_{ij}|\bm{B},\bm{U},\bm{V} \sim \text{PG}\left[k-1, \bm{z}_{ij}^T\text{vec}\left(\bm{B}\right) + \left(\bm{e}_j^{(m)} \otimes \bm{u}_i\right)^T\text{vec}\left(\bm{V}\right)\right].
\end{equation}
Equivalent to the non-latent variable case described above, $\pi\left[\text{vec}\left(\bm{B}\right)\right] = N_{pq}\left(\bm{\mu}_{\text{vec}\left(\bm{B}\right)}^0, \bm{\Sigma}_{\text{vec}\left(\bm{B}\right)}^0\right)$ is assumed, such that the resulting conditional posterior of $\text{vec}\left(\bm{B}\right)$ is given by
\begin{equation}
\text{vec}\left(\bm{B}\right)| \bm{R}, \bm{U}, \bm{V}, \bm{\Omega} \sim N_{pq}\left(\bm{\mu}_{\text{vec}\left(\bm{B}\right)}, \bm{\Sigma}_{\text{vec}\left(\bm{B}\right)}\right),
\end{equation}
where
\begin{equation*}
\bm{\Sigma}_{\text{vec}\left(\bm{B}\right)} = \left[\left(\bm{\Sigma}_{\text{vec}\left(\bm{B}\right)}^0\right)^{-1} + \bm{Z}^T\bm{\Omega}\bm{Z}\right]^{-1}
\end{equation*}
and
\begin{equation*}
\bm{\mu}_{\text{vec}\left(\bm{B}\right)} = \bm{\Sigma}_{\text{vec}\left(\bm{B}\right)}\left\{\left(\bm{\Sigma}_{\text{vec}\left(\bm{B}\right)}^0\right)^{-1}\bm{\mu}_{\text{vec}\left(\bm{B}\right)}^0 + \bm{Z}^T\bm{\Omega}\left[\text{vec}\left(\bm{\Xi}\right) - \left(\bm{I}_m\otimes \bm{U}\right)\text{vec}\left(\bm{V}\right)\right]\right\}.
\end{equation*}
To enforce sparsity on the latent factors, $\pi\left[\text{vec}(\bm{U})\right] = N_{nl}
\left(\bm{0}, \bm{\Sigma}_{\text{vec}(\bm{U})}^0\right)$ is assumed, where $\bm{\Sigma}_{\text{vec}(\bm{U})}^0$ is a $nl \times nl$ diagonal matrix with diagonal values close to 0 for the sparse elements in $\text{vec}(\bm{U})$. A “simple" version of the Bayesian horseshoe model (Carvalho et al., \citeyear{carvalho2009handling}; Makalic \& Schmidt, \citeyear{makalic2015simple}) is used to obtain $\bm{\Sigma}_{\text{vec}(\bm{U})}^0$, as well as $\bm{\Sigma}_{\text{vec}(\bm{V})}^0$, where details are provided in the supplementary material. The simple Bayesian horseshoe model is chosen to ensure that parameter estimation remains free of hyperparameter tuning. Consequently, the resulting conditional posterior of $\text{vec}(\bm{U})$ is given by
\begin{equation}
\text{vec}(\bm{U})| \bm{R}, \bm{B}, \bm{V}, \bm{\Omega} \sim N_{nl}\left[\bm{\mu}_{\text{vec}(\bm{U})}, \bm{\Sigma}_{\text{vec}(\bm{U})}\right],
\end{equation}
where
\begin{equation*}
\bm{\Sigma}_{\text{vec}(\bm{U})} = \left[\left(\bm{\Sigma}_{\text{vec}(\bm{U})}^0\right)^{-1} + (\bm{V} \otimes \bm{I}_n)^T\bm{\Omega}(\bm{V} \otimes \bm{I}_n)\right]^{-1}
\end{equation*}
and
\begin{equation*}
\bm{\mu}_{\text{vec}(\bm{U})} = \bm{\Sigma}_{\text{vec}(\bm{U})}\left\{(\bm{V} \otimes \bm{I}_n)^T\bm{\Omega}\left[\text{vec}\left(\bm{\Xi}\right) - \bm{Z}\text{vec}\left(\bm{B}\right)\right]\right\}.
\end{equation*}
Similarly, $\pi\left[\text{vec}(\bm{V})\right] = N_{ml}\left(\bm{0}, \bm{\Sigma}_{\text{vec}(\bm{V})}^0\right)$ is assumed, such that
\begin{equation}
\text{vec}(\bm{V})| \bm{R}, \bm{B}, \bm{U}, \bm{\Omega} \sim N_{ml}\left[\bm{\mu}_{\text{vec}(\bm{V})}, \bm{\Sigma}_{\text{vec}(\bm{V})}\right],
\end{equation}
where
\begin{equation*}
\bm{\Sigma}_{\text{vec}(\bm{V})} = \left[\left(\bm{\Sigma}_{\text{vec}(\bm{V})}^0\right)^{-1} + (\bm{I}_m \otimes \bm{U})^T\bm{\Omega}(\bm{I}_m \otimes \bm{U})\right]^{-1}
\end{equation*}
and
\begin{equation*}
\bm{\mu}_{\text{vec}(\bm{V})} = \bm{\Sigma}_{\text{vec}(\bm{V})}\left\{(\bm{I}_m \otimes \bm{U})^T\bm{\Omega}\left[\text{vec}\left(\bm{\Xi}\right) - \bm{Z}\text{vec}\left(\bm{B}\right)\right]\right\}.
\end{equation*}
Following a Gibbs sampling procedure similar to the one described in Algorithm \ref{alg:polgam}, $\bm{\Omega}, \bm{B}, \bm{U}$ and $\bm{V}$ are estimated sequentially. Additionally, the latent variable model also allows for computational savings by removing the entries corresponding to any unit-item combination $i$,$j$ for which $r_{ij} = 0$ from $\bm{Z}$, $\bm{\kappa}$, $\bm{\Omega}$, $(\bm{I}_m \otimes \bm{U})$ and $(\bm{V} \otimes \bm{I}_n)$.
\\
\\
The Bayesian horseshoe prior used to obtain $\bm{U}$ and $\bm{F}$ are imposed on $\bm{B}$ when sparse coefficients are assumed. This prior trivializes parameter estimation of sparse $\bm{B}$, as the posterior takes the same form as the one described in Section \ref{Pólya-Gamma data augmentation}, except that $\bm{\Sigma}_0$ is derived in the same way as $\bm{\Sigma}_{\text{vec}(\bm{U})}^0$ and $\bm{\Sigma}_{\text{vec}(\bm{V})}^0$ are for the latent factor model (see the present Section), as is described in the supplementary material. Therefore, further details are omitted.

\section{Simulation study} \label{Simulation study}
This section illustrates the model performance for both the linear, bilinear and latent factor models across a variety of settings. 
\\
\\
The data is simulated in the following manner: for $n \in \{25,50,100,250\}, m \in \{25,50,100,250\}, p,q\in \{5,10\}$ (vectorized) covariate matrices $\text{vec}\left(\bm{X}\right) \sim N_{np}(\bm{0},\bm{I}_{np})$ and $\text{vec}\left(\bm{Y}\right) \sim N_{mq}(\bm{0},\bm{I}_{mq})$ are created. Subsequently, the (vectorised) coefficients are sampled using $\bm{b} \sim N_{p+q}(\bm{0},\bm{I}_{p+q})$ and $\text{vec}\left(\bm{B}\right) \sim N_{pq}(\bm{0},\bm{I}_{pq})$ for the linear and bilinear models respectively. For the latent variable model, the (vectorized) latent factors are sampled as $\text{vec}\left(\bm{U}\right) \sim N_{nl}(\bm{0},\bm{I}_{nl})$ and $\text{vec}\left(\bm{V}\right) \sim N_{ml}(\bm{0},\bm{I}_{ml})$ for $l \in \{1,2\}$, where 75\% of the elements in $\bm{U}$ and $\bm{V}$ are set to 0 to enforce sparsity. Subsequently, either Equation (\ref{eq:eta_def}) or (\ref{eq:eta_def_lat}), depending on whether latent factors are used, is plugged into Equation (\ref{eq:probrij}) to sample the rating data $\bm{R}$, where $k \in \{5,10\}$. However, as real commercial and scientific data frequently contain many unobserved ratings, most of the elements in $\bm{R}$ should be set to zero to resemble real-world scenarios. To elaborate on this, for each user $i$, $|\mathcal{M}_i| \in \{1,5,10\}$ elements in $\mathcal{M}_i$ are uniformly sampled from $\{1,2,\ldots,m\}$. Following this, for $\bm{r}_i, 1\leq i \leq n$, the elements in $\mathcal{M}_i$ are left unchanged, whilst those in $\mathcal{M}_i^c$ are set to 0. To account for the sampling variability of the data, for each combination of parameters, 20 different datasets are generated.
\\
\\
With the data created, the models can be fitted. For each parameter combination, the proposed method is fitted on the 20 simulated datasets using a total of 2000 iterations of the sampler with a burn-in of 1000. Model performance is evaluated using the root mean square error (RMSE), where in addition to parameter estimation, predictive accuracy for the missing elements in $\bm{R}$ is evaluated. The mean was used to obtain a point estimate from the posterior (predictive) distributions. Simulation results without the latent variables are provided in Table \ref{tab:simres1}, whilst the results for the latent variable model are provided in Table \ref{tab:simres2}. Whilst Table \ref{tab:simres1} only contains the results with $|\mathcal{M}_i| = 1$ and $k = 5$, the supplementary material contain results for $|\mathcal{M}_i| \in \{5,10\}$ and $k = 10$. Similarly, Table \ref{tab:simres2} only contains the results for $l = 1$ and $k = 5$, where the other results are shown in the supplementary material.

\begin{table}[H]
%\resizebox{\textwidth}{!}{%
\centering
  \begin{threeparttable}
  \caption{Results for the proposed method on simulated data without latent variables with $|\mathcal{M}_i| = 1$ and $k = 5$. The RMSE is averaged across 20 fitted models for each parameter combination, where the posterior mean was used as a point estimate. Standard deviations are provided between brackets.}
  \label{tab:simres1}
     \begin{tabular}{c | cc | cc }
        \toprule
        \midrule
         & \multicolumn{2}{c|}{Linear} & \multicolumn{2}{c}{Bilinear}\\ \midrule
         \textbf{$n, m, p$} & \textbf{RMSE} $\bm{b}$ & \textbf{RMSE} $\bm{R}$ & \textbf{RMSE} $\bm{B}$ & \textbf{RMSE} $\bm{R}$\\ \midrule
$25, 25, 5$ & 0.43 (0.12) & 0.91 (0.09) & 0.71 (0.12) & 1.27 (0.14)\\ 
$25, 50, 5$ & 0.38 (0.15) & 0.90 (0.13) & 0.63 (0.10) & 1.19 (0.12)\\ 
$25, 100, 5$ & 0.41 (0.14) & 0.92 (0.12) & 0.67 (0.10) & 1.24 (0.10)\\
$25, 250, 5$ & 0.40 (0.13) & 0.88 (0.11) & 0.71 (0.14) & 1.26 (0.15)\\ 
$25, 25, 10$ & 0.59 (0.13) & 1.10 (0.15) & 0.91 (0.08) & 1.67 (0.06)\\ 
$25, 50, 10$ & 0.52 (0.14) & 1.06 (0.11) & 0.92 (0.07) & 1.67 (0.07)\\ 
$25, 100, 10$ & 0.62 (0.15) & 1.11 (0.14) & 0.90 (0.07) & 1.68 (0.06)\\
$25, 250, 10$ & 0.57 (0.15) & 1.11 (0.12) & 0.91 (0.07) & 1.69 (0.04)\\ 
$50, 25, 5$ & 0.23 (0.06) & 0.75 (0.07) & 0.45 (0.10) & 0.97 (0.10)\\
$50, 50, 5$ & 0.28 (0.10) & 0.77 (0.06) & 0.44 (0.11) & 0.97 (0.11)\\ 
$50, 100, 5$ & 0.27 (0.07) & 0.80 (0.09) & 0.45 (0.09) & 0.99 (0.09)\\ 
$50, 250, 5$ & 0.26 (0.08) & 0.81 (0.11) & 0.45 (0.08) & 0.97 (0.07)\\ 
$50, 25, 10$ & 0.38 (0.11) & 0.87 (0.10) & 0.84 (0.06) & 1.54 (0.08)\\ 
$50, 50, 10$ & 0.31 (0.07) & 0.83 (0.05) & 0.81 (0.06) & 1.55 (0.06)\\
$50, 100, 10$ & 0.40 (0.11) & 0.90 (0.10) & 0.83 (0.05) & 1.58 (0.06)\\ 
$50, 250, 10$ & 0.40 (0.10) & 0.89 (0.11) & 0.84 (0.06) & 1.59 (0.04)\\ 
$100, 25, 5$ & 0.20 (0.11) & 0.74 (0.05) & 0.34 (0.08) & 0.81 (0.06)\\ 
$100, 50, 5$ & 0.17 (0.04) & 0.76 (0.07) & 0.28 (0.07) & 0.79 (0.06)\\ 
$100, 100, 5$ & 0.16 (0.04) & 0.76 (0.05) & 0.25 (0.05) & 0.80 (0.05)\\ 
$100, 250, 5$ & 0.17 (0.06) & 0.75 (0.08) & 0.27 (0.06) & 0.79 (0.05)\\ 
$100, 25, 10$ & 0.25 (0.07) & 0.69 (0.08) & 0.69 (0.05) & 1.31 (0.04)\\
$100, 50, 10$ & 0.23 (0.05) & 0.71 (0.07) & 0.69 (0.04) & 1.36 (0.05)\\
$100, 100, 10$ & 0.23 (0.05) & 0.72 (0.04) & 0.66 (0.06) & 1.34 (0.05)\\
$100, 250, 10$ & 0.21 (0.05) & 0.71 (0.04) & 0.66 (0.05) & 1.40 (0.06)\\ 
$250, 25, 5$ & 0.12 (0.04) & 0.68 (0.07) & 0.17 (0.04) & 0.67 (0.04)\\
$250, 50, 5$ & 0.11 (0.03) & 0.70 (0.07) & 0.16 (0.04) & 0.67 (0.03)\\ 
$250, 100, 5$ & 0.10 (0.03) & 0.72 (0.09) & 0.17 (0.04) & 0.67 (0.05)\\
$250, 250, 5$ & 0.10 (0.02) & 0.68 (0.06) & 0.15 (0.03) & 0.67 (0.02)\\  
$250, 25, 10$ & 0.16 (0.07) & 0.63 (0.05) & 0.43 (0.07) & 0.94 (0.06)\\ 
$250, 50, 10$ & 0.15 (0.03) & 0.62 (0.04) & 0.39 (0.06) & 0.94 (0.06)\\ 
$250, 100, 10$ & 0.12 (0.02) & 0.63 (0.05) & 0.37 (0.05) & 0.94 (0.07)\\ 
$250, 250, 10$ & 0.14 (0.03) & 0.64 (0.05) & 0.36 (0.05) & 0.94 (0.06)\\ 
       \midrule
        \bottomrule
     \end{tabular}
  \end{threeparttable}
  %}
\end{table}

\begin{table}[H]
\resizebox{\textwidth}{!}{%
\centering
  \begin{threeparttable}
  \caption{Results for the proposed method on simulated data with latent variables with $l = 1$, $k = 5$ and where the functional form of the predictor is linear. The RMSE is averaged across 20 fitted models for each parameter combination, where the posterior mean was used as a point estimate. Standard deviations are provided between brackets.}
  \label{tab:simres2}
     \begin{tabular}{c | ccc | ccc | ccc }
        \toprule
        \midrule
         & \multicolumn{3}{c|}{$|\mathcal{M}_i| = 1$} & \multicolumn{3}{c}{$|\mathcal{M}_i| = 5$} & \multicolumn{3}{c}{$|\mathcal{M}_i| = 10$}\\ \midrule
         \textbf{$n, m, p$} & \textbf{RMSE} $\bm{b}$ & \textbf{RMSE} $\bm{F}$ & \textbf{RMSE} $\bm{R}$  & \textbf{RMSE} $\bm{b}$ & \textbf{RMSE} $\bm{F}$ & \textbf{RMSE} $\bm{R}$ & \textbf{RMSE} $\bm{b}$ & \textbf{RMSE} $\bm{F}$ & \textbf{RMSE} $\bm{R}$\\ \midrule
$25, 25, 5$ & 0.44 (0.15) & 80.89 (338.40) & 0.94 (0.09) & 0.17 (0.05) & 0.40 (0.31) & 0.73 (0.09) & 0.11 (0.04) & 0.35 (0.21) & 0.71 (0.09)\\ 
$25, 50, 5$ & 0.37 (0.12) & 12.09 (22.60) & 0.90 (0.10) & 0.18 (0.04) & 1.78 (4.36) & 0.74 (0.06) & 0.11 (0.03) & 0.37 (0.33) & 0.71 (0.05)\\ 
$25, 100, 5$ & 0.38 (0.11) & 6.60 (5.47) & 0.91 (0.08) & 0.17 (0.06) & 1.54 (2.79) & 0.74 (0.06) & 0.11 (0.03) & 1.02 (1.06) & 0.72 (0.06)\\ 
$25, 250, 5$ &0.46 (0.12) & 71.10 (139.91) & 0.93 (0.10) & 0.17 (0.06) & 3.11 (5.59) & 0.72 (0.08) &0.13 (0.05) & 2.58 (6.46) & 0.70 (0.08)\\ 
$25, 25, 10$ & 0.58 (0.12) & 45.12 (152.88) & 1.12 (0.14) & 0.24 (0.05) & 0.82 (0.86) & 0.69 (0.06) & 0.17 (0.04) & 0.30 (0.14) & 0.65 (0.06)\\ 
$25, 50, 10$ & 0.62 (0.13) & 65.50 (181.74) & 1.11 (0.15) & 0.22 (0.06) & 1.36 (2.70) & 0.73 (0.06) & 0.15 (0.03) & 0.70 (1.64) & 0.67 (0.07)\\ 
$25, 100, 10$ & 0.65 (0.15) & 81.84 (241.96) & 1.16 (0.09) & 0.27 (0.07) & 1.43 (1.66) & 0.70 (0.05) & 0.15 (0.03) & 0.88 (1.64) & 0.62 (0.05)\\ 
$25, 250, 10$ & 0.62 (0.12) & 130.98 (225.13) & 1.18 (0.10) & 0.24 (0.06) & 6.04 (14.57) & 0.70 (0.05) & 0.17 (0.03) & 1.85 (3.06) & 0.64 (0.05)\\ 
$50, 25, 5$ & 0.27 (0.08) & 14.75 (48.59) & 0.80 (0.06) & 0.11 (0.03) & 0.53 (0.52) & 0.71 (0.06) & 0.08 (0.02) & 0.31 (0.09) & 0.70 (0.06)\\ 
$50, 50, 5$ & 0.25 (0.06) & 11.95 (31.50) & 0.78 (0.05) & 0.11 (0.04) & 0.44 (0.57) & 0.70 (0.05) & 0.08 (0.03) & 0.82 (1.84) & 0.69 (0.04)\\ 
$50, 100, 5$ & 0.26 (0.07) & 5.30 (9.19) & 0.81 (0.07) & 0.11 (0.05) & 0.60 (0.45) & 0.72 (0.07) & 0.08 (0.02) & 0.48 (0.47) & 0.71 (0.07)\\ 
$50, 250, 5$ & 0.28 (0.04) & 56.87 (95.83) & 0.81 (0.08) & 0.13 (0.04) & 1.06 (1.13) & 0.71 (0.04) & 0.09 (0.01) & 0.34 (0.15) & 0.70 (0.03)\\ 
$50, 25, 10$ & 0.38 (0.11) & 4.43 (4.45) & 0.89 (0.10) & 0.16 (0.04) & 0.38 (0.39) & 0.66 (0.05) & 0.10 (0.03) & 0.27 (0.14) & 0.63 (0.06)\\ 
$50, 50, 10$ & 0.34 (0.08) & 6.84 (12.00) & 0.86 (0.08) & 0.13 (0.03) & 0.48 (0.38) & 0.65 (0.04) & 0.10 (0.02) & 0.33 (0.17) & 0.63 (0.04)\\ 
$50, 100, 10$ & 0.39 (0.10) & 5.87 (6.00) & 0.90 (0.09) & 0.14 (0.02) & 0.51 (0.40) & 0.64 (0.04) & 0.10 (0.03) & 0.38 (0.20) & 0.61 (0.04)\\ 
$50, 250, 10$ & 0.35 (0.07) & 8.02 (14.48) & 0.88 (0.06) & 0.14 (0.04) & 1.29 (0.66) & 0.64 (0.02) & 0.10 (0.03) & 0.55 (0.28) & 0.62 (0.03)\\ 
$100, 25, 5$ & 0.18 (0.05) & 1.03 (1.37) & 0.75 (0.06) & 0.07 (0.02) & 0.62 (1.36) & 0.70 (0.06) & 0.06 (0.02) & 0.30 (0.09) & 0.70 (0.06)\\ 
$100, 50, 5$ & 0.17 (0.05) & 2.70 (4.10) & 0.77 (0.07) & 0.08 (0.03) & 0.70 (1.47) & 0.73 (0.07) & 0.05 (0.02) & 0.39 (0.55) & 0.72 (0.07)\\ 
$100, 100, 5$ & 0.18 (0.07) & 7.44 (12.25) & 0.77 (0.05) & 0.07 (0.02) & 0.40 (0.30) & 0.73 (0.04) & 0.05 (0.02) & 0.37 (0.31) & 0.73 (0.05)\\ 
$100, 250, 5$ & 0.21 (0.03) & 3.58 (3.85) & 0.74 (0.06) & 0.08 (0.02) & 1.77 (2.73) & 0.68 (0.06) & 0.05 (0.02) & 0.49 (0.24) & 0.68 (0.06)\\ 
$100, 25, 10$ & 0.30 (0.06) & 3.44 (9.08) & 0.73 (0.05) & 0.12 (0.03) & 0.26 (0.08) & 0.60 (0.05) & 0.08 (0.02) & 0.31 (0.29) & 0.58 (0.05)\\ 
$100, 50, 10$ & 0.25 (0.06) & 3.68 (4.49) & 0.74 (0.06) & 0.10 (0.03) & 0.32 (0.21) & 0.63 (0.05) & 0.07 (0.02) & 0.26 (0.07) & 0.61 (0.05)\\ 
$100, 100, 10$ & 0.27 (0.06) & 3.56 (6.32) & 0.74 (0.07) & 0.10 (0.01) & 1.11 (2.07) & 0.62 (0.03) & 0.07 (0.02) & 0.41 (0.36) & 0.60 (0.04)\\ 
$100, 250, 10$ & 0.25 (0.04) & 20.45 (16.54) & 0.78 (0.07) & 0.09 (0.03) & 0.55 (0.29) & 0.63 (0.02) & 0.06 (0.01) & 0.39 (0.17) & 0.62 (0.03)
\\ 
$250, 25, 5$ & 0.11 (0.04) & 0.87 (1.70) & 0.69 (0.07) & 0.05 (0.02) & 0.32 (0.21) & 0.67 (0.07) & 0.04 (0.01) & 0.27 (0.08) & 0.67 (0.08)\\ 
$250, 50, 5$ & 0.11 (0.04) & 1.07 (1.01) & 0.71 (0.07) & 0.05 (0.01) & 0.53 (0.44) & 0.70 (0.07) & 0.04 (0.01) & 0.32 (0.15) & 0.70 (0.07)\\ 
$250, 100, 5$ & 0.11 (0.03) & 3.24 (3.32) & 0.69 (0.05) & 0.05 (0.01) & 0.59 (0.61) & 0.68 (0.05) & 0.03 (0.01) & 0.30 (0.09) & 0.67 (0.05)\\ 
$250, 250, 5$ & 0.11 (0.01) & 5.11 (8.93) & 0.69 (0.04) & 0.05 (0.02) & 0.40 (0.17) & 0.67 (0.03) & 0.03 (0.01) & 0.29 (0.06) & 0.66 (0.03)\\ 
$250, 25, 10$ & 0.15 (0.04) & 1.52 (2.88) & 0.65 (0.04) & 0.07 (0.01) & 0.36 (0.25) & 0.61 (0.05) & 0.04 (0.01) & 0.30 (0.13) & 0.60 (0.05)\\ 
$250, 50, 10$ & 0.15 (0.03) & 0.88 (0.90) & 0.64 (0.05) & 0.07 (0.02) & 0.32 (0.17) & 0.59 (0.05) & 0.04 (0.01) & 0.29 (0.14) & 0.59 (0.05)\\ 
$250, 100, 10$ & 0.15 (0.05) & 1.77 (2.55) & 0.65 (0.04) & 0.05 (0.01) & 0.27 (0.12) & 0.60 (0.03) & 0.04 (0.01) & 0.23 (0.07) & 0.59 (0.03)\\ 
$250, 250, 10$ & 0.15 (0.04) & 6.48 (10.21) & 0.64 (0.02) & 0.07 (0.02) & 1.28 (1.64) & 0.59 (0.03) & 0.05 (0.01) & 0.25 (0.06) & 0.59 (0.03)\\ 
       \midrule
        \bottomrule
     \end{tabular}
  \end{threeparttable}}
  %}
\end{table}

\noindent In general, judging from Tables \ref{tab:simres1} and \ref{tab:simres2},  the proposed method is able to accurately estimate the parameters, especially for large values of $n$ or $\left|\mathcal{M}_i\right|$. Note, however, that an increase in $m$ does not improve parameter estimation, as the number of observed ratings does not increase in $m$. Even though the parameters can be estimated with arbitrary accuracy, this does not hold for the predictive accuracy. This limit is an artefact of the probabilistic nature of the shifted binomial model. As such, the posterior predictive distributions are multimodal for $r_{ij}$ whenever the corresponding $p_{ij}$, see Equation (\ref{eq:probrij}), is not close to either 0 or 1. 

\subsection{Competing with recommender systems}
Whereas the proposed method performs well on the model-based simulations shown above, its performance is contingent upon untestable parametric assumptions. Whether or not this poses a problem for real-world data is something that merits evaluation. To this end, another simulation study is conducted, where four different real-world datasets are used to asses the predictive capability of the proposed method and all its extensions against five commonly used recommender systems. These datasets are the Amazon music dataset (Hou et al., \citeyear{hou2024bridging}), the Goodbooks dataset (Zajac, \citeyear{goodbooks2017}), the MovieLens dataset (Harper \& Konstan, \citeyear{harper2015movielens}) and the Yelp dataset (Yelp, \citeyear{yelp2024}). Instead of using all four datasets in their entirety, for each dataset the ratings on the top 100 items by the top 100 users are chosen, where each user is required to have rated at least one of the top 100 items. This results in 1319, 3306, 7086 and 1678 observed ratings for the Amazon music, Goodbooks, MovieLens and Yelp datasets respectively. Both user and item covariates are available for all datasets, and are utilized by the proposed method. Details on these covariates are provided in the supplementary material. The recommender systems against which the proposed method is evaluated are: the singular value decomposition (SVD) (Koren et al., \citeyear{koren2009matrix}), Simon Funk's singular value decomposition (Funk SVD) (Koren et al., \citeyear{koren2009matrix}), alternating least squares (ALS) (Koren et al., \citeyear{koren2009matrix}), item-based collaborative filtering (IBCF) (Kitts et al., \citeyear{kitts2000cross}) and user-based collaborative filtering (UBCF) (Goldberg et al., \citeyear{goldberg1992using}).
\\
\\
\noindent This simulation study proceeds as follows: using a five-fold cross-validation scheme, each dataset is divided into five folds, where each fold alternatively makes up the testing data and the others make up the training data. Each of the models is fitted on the training data, and the ratings in the testing data are predicted using the fitted models. As these ratings are known, the RMSE is computed on these values. This is done for all five folds, after which the average of these five RMSE values is computed. The results are provided in Table \ref{tab:reccomp}.

\begin{table}[H]
%\resizebox{\textwidth}{!}{%
\centering
  \begin{threeparttable}
  \caption{Predictive accuracy results for various methods on four real-world datasets. ALS stands for alternating least squares, IBCF for item-based collaborative filtering, UBCF for user-based collaborative filtering, Lin for the proposed method with a linear predictor and Bilin for the proposed method with a bilinear predictor. The RMSE on the predicted ratings is averaged across five folds, where the posterior mean was used as a point estimate. Standard deviations are provided between brackets.}
  \label{tab:reccomp}
     \begin{tabular}{c | c | c | c | c }
        \toprule
        \midrule
              \diagbox[width=\dimexpr \textwidth/8+2\tabcolsep\relax]{Method}{Data}
 & \multicolumn{1}{c|}{Amazon music} & \multicolumn{1}{c|}{Goodbooks} & \multicolumn{1}{c|}{MovieLens} & \multicolumn{1}{c}{Yelp}\\ \midrule
SVD & 1.06 (0.08) & \textbf{0.89 (0.03)} & 0.90 (0.02) & 0.93 (0.04)\\ 
Funk SVD & 1.04 (0.05) & 0.91 (0.04) & 0.88 (0.02) & 0.92 (0.04)\\
ALS & 1.15 (0.06) & 0.92 (0.01) & 0.90 (0.02) & 1.05 (0.04)\\
IBCF & 1.14 (0.08) & 0.91 (0.03) & 0.89 (0.02) & 1.02 (0.03)\\
UBCF & 1.14 (0.04) & 0.91 (0.04) & 0.89 (0.03) & 0.99 (0.04)\\
Lin & 1.08 (0.06) & 0.91 (0.04) & 0.90 (0.02) & 0.87 (0.03)\\
Lin sparse & 1.08 (0.06) & 0.91 (0.04) & 0.90 (0.03) & 0.87 (0.03)\\
Bilin & 1.12 (0.07) & 0.91 (0.05) & 0.89 (0.03) & 0.95 (0.05)\\
Bilin sparse & 1.12 (0.06) & 0.91 (0.04) & 0.88 (0.03) & 0.91 (0.04)\\
Lin $l=1$ & \textbf{1.03 (0.07)} & 0.91 (0.04) & 0.88 (0.03) & \textbf{0.87 (0.03)}\\
Lin $l=2$ & 1.05 (0.07) & 0.91 (0.04) & 0.87 (0.03) & 0.87 (0.03)\\
Lin sparse $l=1$ & 1.06 (0.05) & 0.90 (0.04) & 0.88 (0.03) & 0.87 (0.03)\\
Lin sparse $l=2$ & 1.10 (0.05) & 0.90 (0.04) & \textbf{0.87 (0.03)} & 0.87 (0.03)\\
Bilin $l=1$ & 1.11 (0.12) & 0.91 (0.04) & 0.88 (0.03) & 0.96 (0.04)\\
Bilin $l=2$ & 1.10 (0.05) & 0.91 (0.03) & 0.89 (0.02) & 0.95 (0.04)\\
Bilin sparse $l=1$ & 1.08 (0.10) & 0.91 (0.04) & 0.87 (0.03) & 0.92 (0.04)\\
Bilin sparse $l=2$ & 1.08 (0.05) & 0.91 (0.05) & 0.88 (0.03) & 0.91 (0.04)\\
       \midrule
        \bottomrule
     \end{tabular}
  \end{threeparttable}
 % }
\end{table}

\noindent Judging from the results provided by Table \ref{tab:reccomp}, the proposed method is able to hold its own against commonly used recommender systems in prediction tasks. Especially the linear models with latent factors, both the sparse and non-sparse versions, appear capable of striking a a balance between complexity and simplicity, resulting in relatively accurate predictions. Moreover, the worst performing version of the proposed method outperforms the worst performing recommender system on each of the four datasets. Whilst the differences between the best and worst performing methods are relatively small for the Goodbooks and Movielens datasets, substantial differences between these two extremes are observed in the Amazon music and Yelp data. To conclude, it appears that focusing on uncertainty quantification by making parametric assumptions on the underlying data generating process does not necessarily result in decreased predictive capabilities.

\section{A statistical matchmaker} \label{Real data application}
\begin{figure}[H]
\centering
\includegraphics[width=0.9\textwidth]{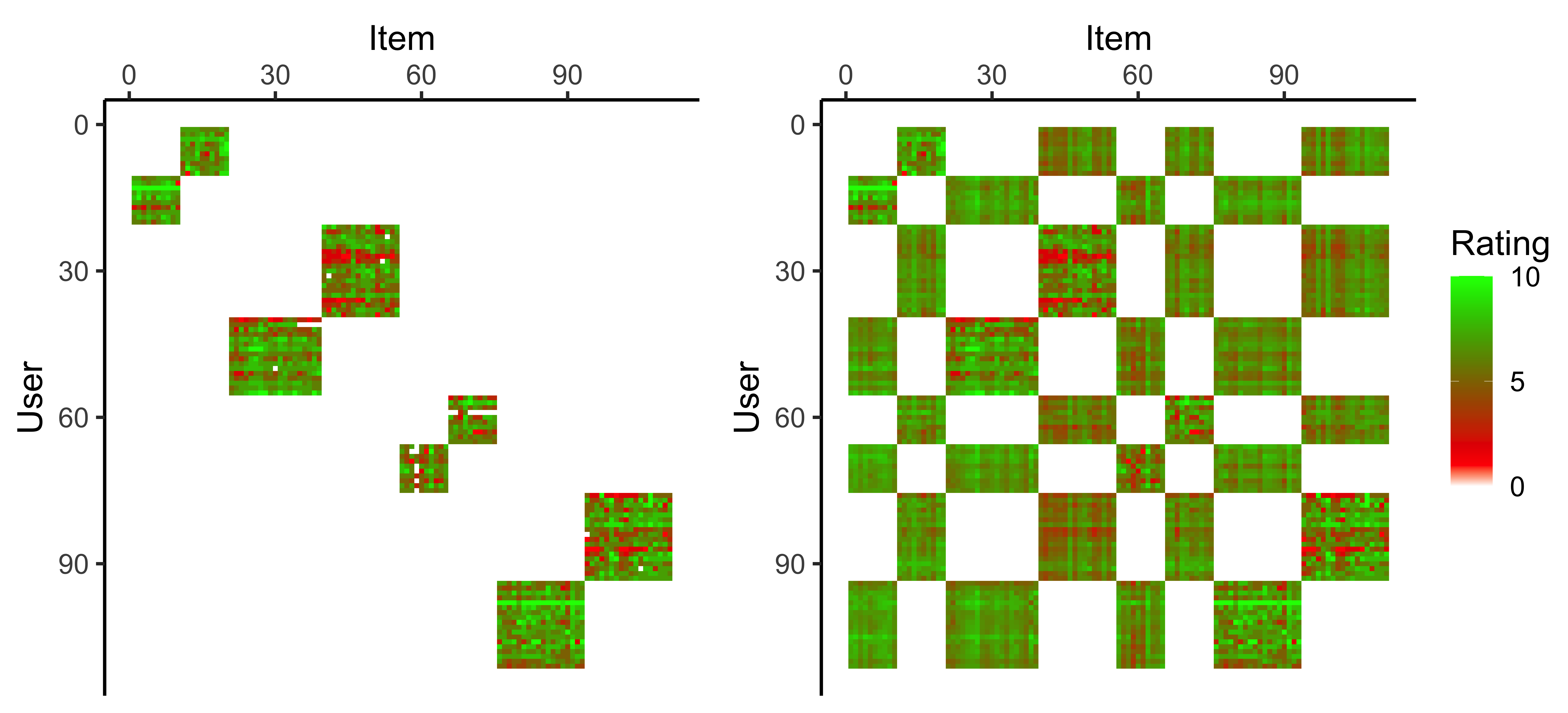}
\caption{Observed ratings for overall liking (left) and predicted ratings (right) from four rounds of speed dating. The predicted ratings are based on the mean of the posterior predictive distributions. Only those predicted ratings contained within the set of heterosexual pairings are shown, adhering to the sexual preferences of the students.}
\label{fig:dating_ratings}
\end{figure}

Finding the right partner in life is not always easy, and depends, amongst others, on characteristics related to the individual looking for love, those of the potential partner, and the between-individual interaction of these characteristics (Hitsch et al., \citeyear{hitsch2010makes}). Whilst having been romanticised by various books and movies, economists aim to study this problem in a scientific manner (Sprecher et al., \citeyear{sprecher1991effect}; Fisman et al., \citeyear{fisman2008racial}; Hitsch et al., \citeyear{hitsch2010matching}; \citeyear{hitsch2010makes}; Bojd \& Yoganarasimhan, \citeyear{bojd2022star}). One way to study this problem is to design an experiment whereby various pairs of individuals are send off on dates, after which they rate the person they dated on various attributes, such as (physical) attractiveness, kindness, etc. The economist then relates these observed ratings to the characteristics of each individual to infer the driving forces of partner preferences (Fisman et al., \citeyear{fisman2008racial}). 
\\
\\
In this section, the proposed method is applied on speed dating data collected by Fisman et al.\ (\citeyear{fisman2008racial}), in order to illustrate the added value of the proposed method on the problem of partner preferences, and the adjacent problem of partner choice. Observing that interracial marriages are an uncommon occurrence in the US, Fisman et al.\ (\citeyear{fisman2008racial}) became interested in examining racial preferences in dating. To evaluate these preferences, they employed a speed dating experiment, where students in graduate and professional schools at Columbia University were seated at two-person tables, at which each student had a conversation with another student of the opposite sex. These conversations lasted for four minutes, after which the participants filled in a form, expressing their impression of the other person, including how much they liked this person overall on a Likert scale from one through ten. Subsequently, the male students moved to the adjacent table, whilst the female students remained seated, such that all students of opposite sexes ended up on a speed date with each other.  
\\
\\
The data used in this analysis contains the first four waves of speed dates found in Fisman et al.\ (\citeyear{fisman2008racial}), with each wave containing different students, and where at each wave the students dated one another, but no dates occurred between students of different waves. These four waves consists of a total of 57 females and 54 males. In addition to the gender of these 111 students, their age and race are known. Moreover, for each student, their average attractiveness, sincerity, intelligence, fun and ambitiousness are computed based on the scores of these qualities provided by their dates. The race variable is turned into five dummy variables: white, black, Hispanic, East Asian and native, where “other” serves as the baseline race.  Furthermore, the data contain a total of 1656 ratings, reflecting how much each student likes their date. Therefore, using the terminology found throughout this article, each student is both user and item. The observed ratings for overall liking are shown in Figure \ref{fig:dating_ratings}. Whilst this rating was not analysed directly by Fisman et al.\ (\citeyear{fisman2008racial}), overall liking is a more complex trait than attractiveness, which Fisman et al.\ (\citeyear{fisman2008racial}) did analyse in their study, and is therefore interesting in the analysis of partner preferences. Moreover, the usage of ordinary least squares (OLS) by Fisman et al.\ (\citeyear{fisman2008racial}) with attractiveness as the dependent variable is curious, given that attractiveness is also rated on a Likert scale from one through ten. Therefore, the proposed method seems better suited for this type of analysis, as OLS assumptions are violated with an ordinal dependent variable, given the non-normality of the residuals, and OLS predictions are not in $\mathcal{K}$. Additionally, an ordinal type model should be preferred over OLS, because the substantive difference between a 4 and 5 star rating might not be the same as the difference between a 3 and 4 star rating, which is what OLS assumes.

\begin{figure}[H]
\centering
\includegraphics[width=0.5\textwidth]{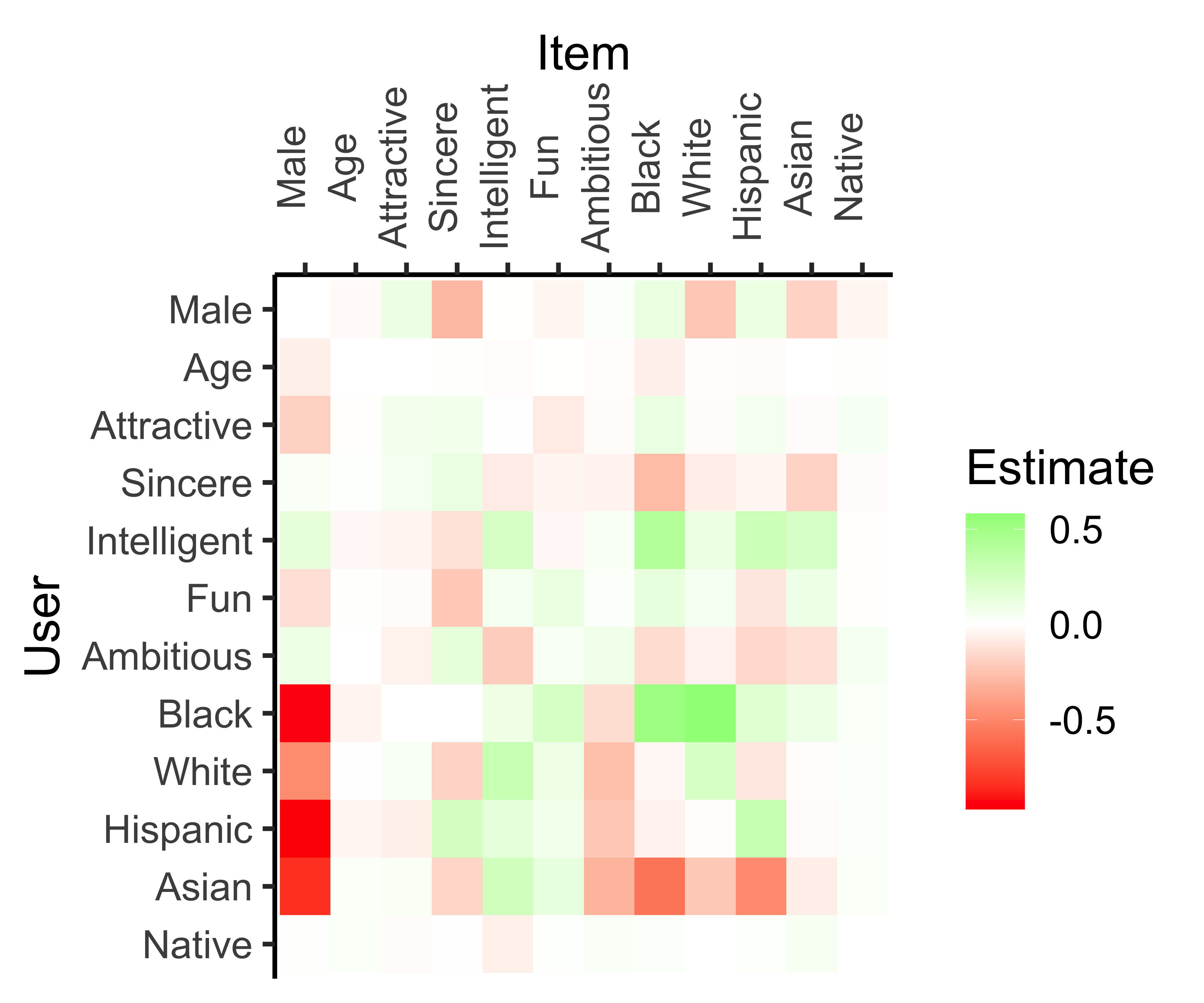}
\caption{Posterior means for the parameter estimates, based on 5000 Gibbs samples.}
\label{fig:postmed}
\end{figure}

\noindent In their analysis, Fisman et al.\ (\citeyear{fisman2008racial}) primarily look at interaction effects between variables such as age, race, gender and attractiveness. To allow for such interactions in this application, the bilinear model is fitted on the data. Rather than repeating the findings of Fisman et al.\ (\citeyear{fisman2008racial}), the goal here is to show that the proposed method can acquire new insights into the partner preferences governing the dating market. A total of $T = 5000$ iterations of Algorithm \ref{alg:polgam} are run, of which the first 2500 served as burn-in. The posterior means for the parameter estimates are provided in Figure \ref{fig:postmed}.
\noindent Whilst several similar results to those of Fisman et al.\ (\citeyear{fisman2008racial}) are obtained, such as the strong same race preferences for black, white and hispanic students, some additional interesting relationships are found, which merit some discussion. One such result is that despite the strong same race preference for black students (posterior mean of 0.52 with 95\% CI [-0.49, 1.53]), they appear to exhibit an even stronger preference for a white partner (posterior mean of 0.58 with 95\% CI [-0.14, 1.27]), as judged from the slightly narrower and positively shifted credibility interval (CI). However, this preference is not observed the other way around (posterior mean of -0.04 with 95\% CI [-0.78, 0.69]). Moreover, female students have a preference for white and asian partners (men have posterior means of respectively -0.23 and -0.18 with 95\% CI [-0.59, 0.12] and [-0.56, 0.22]), compared to male students who have a preference for a black or hispanic partner (posterior means of respectively 0.12 and 0.11 with 95\% CI [-0.38, 0.61] and [-0.31, 0.55]). Another result is that although having an ambitious partner seems not desirable to any of the races relative to the baseline “other" race, having an intelligent and fun partner appears all the more important. A final remarkable result is the very strong preference of intelligent students for a black partner (posterior mean of 0.41 with 95\% CI [-0.14, 0.97]), whilst black partners are least desired by sincere students (posterior mean of -0.28 with 95\% CI [-0.69, 0.14]). 

\section{Conclusion} \label{Conclusion and discussion}
By introducing a novel statistical method, this paper aims to contribute to inferential questions on multi-item rating and recommendation problems. By introducing several extensions, the applicability of the proposed method is increased further. Even though the proposed method is not designed with maximising predictive accuracy in mind, the model is able to hold its own in terms of prediction against commonly used recommender systems across various real-world datasets. Applying the model on data pertaining to a speed dating experiment provides new insights into the romantic preferences of people.  

In terms of future research, the proposed method can be extended to handle ratings on different attributes, e.g. taste, smell, novelty whenever the items are different foods. Another extension can be to include contextual information, or to model ratings together with user-provided reviews.

\bibliographystyle{Chicago}
\bibliography{library}

\section*{Supplementary material}

\subsection*{Identifiability proof}

\noindent Proof of Theorem 1.
\begin{proof}
To simplify notation, the proof is written in terms of $\bm{Z}$ instead of $\bm{X}$ and $\bm{Y}$. Suppose some $\bm{R} \in \{\mathcal{K} \cup \{0\}\}^{n \times m}$ and $\bm{Z} \in \mathbb{R}^{nm \times pq}$ are observed. Suppose further that $\mathbb{P}\left(\bm{R}|\bm{B}_1\right) = \mathbb{P}\left(\bm{R}|\bm{B}_1\right)$, such that
\begin{equation}
\begin{gathered}
\prod_{i = 1}^n\prod_{j \in \mathcal{M}_i}\binom{k - 1}{r_{ij} - 1}\frac{\exp\left[\bm{z}_{ij}^T\text{vec}\left(\bm{B}_1\right)\right]^{r_{ij} - 1}}{\left\{1 + \exp\left[\bm{z}_{ij}^T\text{vec}\left(\bm{B}_1\right)\right]\right\}^{k - 1}} = \prod_{i = 1}^n\prod_{j \in \mathcal{M}_i}\binom{k - 1}{r_{ij} - 1}\frac{\exp\left[\bm{z}_{ij}^T\text{vec}\left(\bm{B}_2\right)\right]^{r_{ij} - 1}}{\left\{1 + \exp\left[\bm{z}_{ij}^T\text{vec}\left(\bm{B}_2\right)\right]\right\}^{k - 1}}\\
%\sum_{i = 1}^n\sum_{j \in \mathcal{M}_i}\log\binom{k - 1}{r_{ij} - 1} + \sum_{i = 1}^n\sum_{j \in \mathcal{M}_i}\left(r_{ij}-1\right)\bm{z}_{ij}^T\text{vec}\left(\bm{B}_1\right) - \sum_{i = 1}^n\sum_{j \in \mathcal{M}_i}\left(k - 1\right)\log\left\{1 + \exp\left[\bm{z}_{ij}^T\text{vec}\left(\bm{B}_1\right)\right]\right\}\\
%- \sum_{i = 1}^n\sum_{j \in \mathcal{M}_i}\log\binom{k - 1}{r_{ij} - 1} - \sum_{i = 1}^n\sum_{j \in \mathcal{M}_i}\left(r_{ij}-1\right)\bm{z}_{ij}^T\text{vec}\left(\bm{B}_2\right) + \sum_{i = 1}^n\sum_{j \in \mathcal{M}_i}\left(k - 1\right)\log\left\{1 + \exp\left[\bm{z}_{ij}^T\text{vec}\left(\bm{B}_2\right)\right]\right\}\\
%=  \sum_{i = 1}^n\sum_{j \in \mathcal{M}_i}\left(k - 1\right)\log\left\{1 + \exp\left[\bm{z}_{ij}^T\text{vec}\left(\bm{B}_2\right)\right]\right\} - \sum_{i = 1}^n\sum_{j \in \mathcal{M}_i}\left(k - 1\right)\log\left\{1 + \exp\left[\bm{z}_{ij}^T\text{vec}\left(\bm{B}_1\right)\right]\right\}\\
%+ \sum_{i = 1}^n\sum_{j \in \mathcal{M}_i}\left(r_{ij}-1\right)\bm{z}_{ij}^T\text{vec}\left(\bm{B}_1\right) -  \sum_{i = 1}^n\sum_{j \in \mathcal{M}_i}\left(r_{ij}-1\right)\bm{z}_{ij}^T\text{vec}\left(\bm{B}_2\right)\\
%\sum_{i = 1}^n\sum_{j \in \mathcal{M}_i}\left(k - 1\right)\log\left\{\frac{1 + \exp\left[\bm{z}_{ij}^T\text{vec}\left(\bm{B}_2\right)\right]}{1 + \exp\left[\bm{z}_{ij}^T\text{vec}\left(\bm{B}_1\right)\right]}\right\} + \sum_{i = 1}^n\sum_{j \in \mathcal{M}_i}\left(r_{ij}-1\right)\bm{z}_{ij}^T\left[\text{vec}\left(\bm{B}_1\right) - \text{vec}\left(\bm{B}_2\right)\right] = 0
\sum_{i = 1}^n\sum_{j \in \mathcal{M}_i}\left\{\left(k - 1\right)\log\left\{\frac{1 + \exp\left[\bm{z}_{ij}^T\text{vec}\left(\bm{B}_2\right)\right]}{1 + \exp\left[\bm{z}_{ij}^T\text{vec}\left(\bm{B}_1\right)\right]}\right\} + \left(r_{ij}-1\right)\bm{z}_{ij}^T\left[\text{vec}\left(\bm{B}_1\right) - \text{vec}\left(\bm{B}_2\right)\right]\right\} = 0.
\end{gathered}
\end{equation}
Above equality holds iff $\bm{Z}\text{vec}\left(\bm{B}_1\right) = \bm{Z}\text{vec}\left(\bm{B}_2\right)$. This is true irrespective of the rank of $\bm{Z}$, whenever $\text{vec}\left(\bm{B}_1\right) = \text{vec}\left(\bm{B}_1\right)$. However, note that if rank$(\bm{Z}) < pq$, then it is well known that for any linear system of equations, there exists $\text{vec}\left(\bm{B}_1\right) \neq \text{vec}\left(\bm{B}_2\right)$ whilst $\bm{Z}\text{vec}\left(\bm{B}_1\right) = \bm{Z}\text{vec}\left(\bm{B}_2\right)$. Therefore, rank$(\bm{Z})$ is required to equal $pq$. Consequently, as $\bm{Z} = \bm{Y} \otimes \bm{X}$, it follows that $\text{rank}\left(\bm{Z}\right) = pq$ iff $\text{rank}\left(\bm{X}\right) = p$ and $\text{rank}\left(\bm{Y}\right) = q$, completing the proof.
\end{proof}

\subsection*{Pólya-gamma derivation}
Below, derivations are provided showing that the full conditional distribution of $\text{vec}\left(\bm{B}\right)$ is equal to Equation (10). To simplify notation, let $\bm{z}_{ij} = \bm{y}_{j} \otimes \bm{x}_{i}$, then
\begin{equation}
\begin{gathered}
\mathbb{P}\left[\text{vec}\left(\bm{B}\right)|\bm{\omega}\right] \propto \pi\left[\text{vec}\left(\bm{B}\right)\right]\prod_{i = 1}^n\prod_{j \in \mathcal{M}_i}\mathbb{P}\left[r_{ij}|\text{vec}\left(\bm{B}\right)\right]\mathbb{P}\left[\omega_{ij}|r_{ij},\text{vec}\left(\bm{B}\right)\right]\\
= \pi\left[\text{vec}\left(\bm{B}\right)\right]\prod_{i = 1}^n\prod_{j \in \mathcal{M}_i}\frac{\exp\left[\bm{z}_{ij}^T\text{vec}\left(\bm{B}\right)\right]^{r_{ij} - 1}}{\left\{1 + \exp\left[\bm{z}_{ij}^T\text{vec}\left(\bm{B}\right)\right]\right\}^{k - 1}}\mathbb{P}\left[\omega_{ij}|r_{ij},\text{vec}\left(\bm{B}\right)\right]\\
= \pi\left[\text{vec}\left(\bm{B}\right)\right]\prod_{i = 1}^n\prod_{j \in \mathcal{M}_i}\left\{\exp\left[\kappa_{ij}\bm{z}_{ij}^T\text{vec}\left(\bm{B}\right)\right]\int_{0}^\infty\exp\left[-\frac{1}{2}\omega_{ij}[\bm{z}_{ij}^T\text{vec}\left(\bm{B}\right)]^2\right]\mathbb{P}\left(\omega_{ij}|k-1,0\right)d\omega_{ij}\right\}\mathbb{P}\left[\omega_{ij}|r_{ij},\text{vec}\left(\bm{B}\right)\right]\\
= \pi\left[\text{vec}\left(\bm{B}\right)\right]\prod_{i = 1}^n\prod_{j \in \mathcal{M}_i}\left\{\exp\left[\kappa_{ij}\bm{z}_{ij}^T\text{vec}\left(\bm{B}\right)\right]\int_{0}^\infty\exp\left[-\frac{1}{2}\omega_{ij}[\bm{z}_{ij}^T\text{vec}\left(\bm{B}\right)]^2\right]\mathbb{P}\left(\omega_{ij}|k-1,0\right)d\omega_{ij}\right\}\\
\times \prod_{i = 1}^n\prod_{j \in \mathcal{M}_i}\frac{\exp[-\frac{1}{2}\omega_{ij}(\bm{z}_{ij}^T\text{vec}\left(\bm{B}\right)]^2]\mathbb{P}(\omega_{ij}|k-1, 0)}{\int_{0}^\infty\exp[-\frac{1}{2}\omega_{ij}[\bm{z}_{ij}^T\bm{\beta})^2]\mathbb{P}(\omega_{ij}|k-1, 0)d\omega_{ij}}\\
= \pi\left[\text{vec}\left(\bm{B}\right)\right]\prod_{i = 1}^n\prod_{j \in \mathcal{M}_i}\exp\left[\kappa_{ij}\bm{z}_{ij}^T\text{vec}\left(\bm{B}\right)\right]\exp\left[-\frac{1}{2}\omega_{ij}[\bm{z}_{ij}^T\text{vec}\left(\bm{B}\right)]^2\right]\mathbb{P}\left(\omega_{ij}|k-1,0\right) \quad \left(\text{integrals cancel out}\right)\\
= \pi\left[\text{vec}\left(\bm{B}\right)\right]\prod_{i = 1}^n\prod_{j \in \mathcal{M}_i}\exp\left[\kappa_{ij}\bm{z}_{ij}^T\text{vec}\left(\bm{B}\right)\right]\exp\left[-\frac{1}{2}\omega_{ij}[\bm{z}_{ij}^T\text{vec}\left(\bm{B}\right)]^2\right] \quad \left(\mathbb{P}\left(\omega_{ij}|k-1,0\right) \text{is constant w.r.t.} \text{vec}\left(\bm{B}\right)\right)\\
= \pi\left[\text{vec}\left(\bm{B}\right)\right]\prod_{i = 1}^n\prod_{j \in \mathcal{M}_i}\exp\left\{\kappa_{ij}\bm{z}_{ij}^T\text{vec}\left(\bm{B}\right) -\frac{1}{2}\omega_{ij}[\bm{z}_{ij}^T\text{vec}\left(\bm{B}\right)]^2\right\}\\
= \pi\left[\text{vec}\left(\bm{B}\right)\right]\prod_{i = 1}^n\prod_{j \in \mathcal{M}_i}\exp\left\{-\frac{1}{2}\omega_{ij}\left[\left[\bm{z}_{ij}^T\text{vec}\left(\bm{B}\right)\right]^2 - 2\left[\bm{z}_{ij}^T\text{vec}\left(\bm{B}\right)\right]\xi_{ij}\right]\right\}\\
= \pi\left[\text{vec}\left(\bm{B}\right)\right]\prod_{i = 1}^n\prod_{j \in \mathcal{M}_i}\exp\left\{-\frac{1}{2}\omega_{ij}\left[\xi_{ij} - \bm{z}_{ij}^T\text{vec}\left(\bm{B}\right)\right]^2\right\}\\
\propto \pi\left[\text{vec}\left(\bm{B}\right)\right]\exp\left\{-\frac{1}{2}\left[\text{vec}\left(\bm{\Xi}\right) - \bm{Z}\text{vec}\left(\bm{B}\right)\right]^T\bm{\Omega}\left[\text{vec}\left(\bm{\Xi}\right) - \bm{Z}\text{vec}\left(\bm{B}\right)\right]\right\}
%\propto \pi\left(\bm{\beta}\right)\exp\left[-\frac{1}{2}\left(\text{vec}\left(\bm{\Xi}\right) - \bm{Z}\text{vec}\left(\bm{B}\right)\right)^T\bm{\Omega}\left(\text{vec}\left(\bm{\Xi}\right) - \bm{Z}\text{vec}\left(\bm{B}\right)\right)\right]
\end{gathered}
\end{equation}
where $\bm{\Xi} = (\xi_{ij})_{n \times m}, \bm{\Omega} = \text{diag}(\omega_{11},\omega_{21},\ldots,\omega_{n1},\ldots,\omega_{1m},\omega_{2m},\ldots,\omega_{nm})$, with $\bm{\Omega} \in \mathbb{R}^{nm\times nm}$, and where $\xi_{ij} = \kappa_{ij}/\omega_{ij}$, with $\omega_{ij} \neq 0 \Leftrightarrow j \in \mathcal{M}_{i}$ and $\kappa_{ij} \neq 0 \Leftrightarrow j \in \mathcal{M}_{i}$. 
\\
\\
Similar results can be obtained for the latent factor model by replacing the predictor $\bm{z}_{ij}^T\text{vec}\left(\bm{B}\right)$ with $\bm{z}_{ij}^T\text{vec}\left(\bm{B}\right) + \left(\bm{e}_j^{(m)} \otimes \bm{u}_i\right)^T\text{vec}\left(\bm{V}\right)$.

\subsection*{Sparse latent factors}
Sparse estimates of $\bm{U}$ and $\bm{V}$ are obtained by embedding the “simple" Bayesian horseshoe scheme described in Makalic and Schmidt (\citeyear{makalic2015simple}) into the Bayesian estimation framework described in Section 3 of the main text. Here, the priors for $\text{vec}\left(\bm{U}\right)$ and $\text{vec}\left(\bm{V}\right)$ are 0-centered multivariate normal distributions, with diagonal covariance matrices $\bm{\Sigma}_{\text{vec}\left(\bm{U}\right)}^0$ and  $\bm{\Sigma}_{\text{vec}\left(\bm{V}\right)}^0$ respectively, whose diagonal entries correspond to 0 elements in the vectorized latent factors are close to 0, which in turn ensures that these elements are also close to 0 in the posterior covariance matrix. The process of obtaining $\bm{\Sigma}_{\text{vec}\left(\bm{U}\right)}^0$ is described below, where the steps for $\bm{\Sigma}_{\text{vec}\left(\bm{V}\right)}^0$ are very similar, except that the quantities corresponding to $\bm{U}$ are replaced by those corresponding to $\bm{V}$, and the indices $1,\ldots,n$ are replaced by $1,\ldots,m$.
\\
\\
The following random variables are sequentially sampled from an inverse-gamma distribution $IG(\cdot,\cdot)$ at each iteration of the Gibbs sampling algorithm
\begin{equation*}
\left(\lambda_{il'}^{(u)}\right)^2|\nu_{il'}^{(u)},\left(\tau^{(u)}\right)^2,u_{il'} \sim IG\left(1, \frac{1}{\nu_{il'}^{(u)}} + \frac{u_{il'}^2}{\left(\tau^{(u)}\right)^2}\right), \quad (i = 1,2,\ldots,n), (l' = 1,2,\ldots,l),
\end{equation*} 

\begin{equation*}
\left(\tau^{(u)}\right)^2|\nu_{il'}^{(u)}, \left(\lambda_{il'}^{(u)}\right)^2, u_{il'},\zeta^{(u)} \sim IG\left(\frac{nl+1}{2}, \frac{1}{\zeta^{(u)}} + \frac{1}{2}\sum_{i = 1}^n\sum_{l' = 1}^l\frac{u_{il'}^2}{\left(\lambda_{il'}^{(u)}\right)^2}\right),
\end{equation*} 

\begin{equation*}
\nu_{il'}^{(u)}|\left(\lambda_{il'}^{(u)}\right)^2 \sim IG\left(1, 1+ \frac{1}{\left(\lambda_{il'}^{(u)}\right)^2}\right), \quad (i = 1,2,\ldots,n), (l' = 1,2,\ldots,l),
\end{equation*} 

\begin{equation*}
\zeta^{(u)}|\left(\tau^{(u)}\right)^2 \sim IG\left(1, 1+ \frac{1}{\left(\tau^{(u)}\right)^2}\right).
\end{equation*} 
Consequently, set $\bm{\Lambda}^{(u)} = \text{diag}\left[\left(\lambda_{11}^{(u)}\right)^2,\left(\lambda_{21}^{(u)}\right)^2,\ldots,\left(\lambda_{nl}^{(u)}\right)^2\right]$, such that $\bm{\Sigma}_{\text{vec}\left(\bm{U}\right)}^0 = \left[\left(\tau^{(u)}\right)^2\bm{\Lambda}^{(u)}\right]^{-1}$ is obtained, which is used in Equation (14).

\subsection{Additional simulation results}
The simulations in main text reflect only a fraction of the possible configurations that can be encountered in practice. Therefore, this section contains some additional results, focussing on larger $\left|\mathcal{M}_i\right|$ for the linear and bilinear models (Table \ref{tab:simressup1}), higher $k$ for the linear and bilinear models (Table \ref{tab:simressup2}), higher $k$ for the latent variable model with $\ell = 1$ (Table \ref{tab:simressup3}), the latent variable model with $\ell = 2$ (Table \ref{tab:simressup4}) and the latent variable model with higher $k$ and $\ell = 2$ (Table \ref{tab:simressup5}).

\begin{table}[H]
\resizebox{\textwidth}{!}{%
\centering
  \begin{threeparttable}
  \caption{Results for the proposed method on simulated data without latent variables with $k = 5$. The RMSE is averaged across 20 fitted models for each parameter combination, where the posterior mean was used as a point estimate. Standard deviations are provided between brackets.}
  \label{tab:simressup1}
     \begin{tabular}{c | cc | cc | cc | cc }
        \toprule
        \midrule
         & \multicolumn{4}{c|}{Linear} & \multicolumn{4}{c}{Bilinear}\\ \midrule
         & \multicolumn{2}{c|}{$|\mathcal{M}_i| = 5$} & \multicolumn{2}{c|}{$|\mathcal{M}_i| = 10$} & \multicolumn{2}{c|}{$|\mathcal{M}_i| = 5$} & \multicolumn{2}{c}{$|\mathcal{M}_i| = 10$}\\ \midrule
         \textbf{$n, m, p$} & \textbf{RMSE} $\bm{b}$ & \textbf{RMSE} $\bm{R}$ & \textbf{RMSE} $\bm{B}$ & \textbf{RMSE} $\bm{R}$ & \textbf{RMSE} $\bm{b}$ & \textbf{RMSE} $\bm{R}$ & \textbf{RMSE} $\bm{B}$ & \textbf{RMSE} $\bm{R}$\\ \midrule
$25, 25, 5$ & 0.16 (0.04) & 0.71 (0.08) & 0.12 (0.04) & 0.70 (0.07) & 0.30 (0.05) & 0.77 (0.06) & 0.18 (0.03) & 0.69 (0.05)\\
$25, 50, 5$ & 0.16 (0.05) & 0.72 (0.06) & 0.12 (0.04) & 0.69 (0.05) & 0.25 (0.05) & 0.74 (0.06) & 0.16 (0.03) & 0.67 (0.04)\\ 
$25, 100, 5$ & 0.15 (0.04) & 0.73 (0.06) & 0.10 (0.03) & 0.70 (0.07) & 0.26 (0.05) & 0.74 (0.04) & 0.18 (0.04) & 0.68 (0.04)\\
$25, 250, 5$ & 0.16 (0.06) & 0.70 (0.07) & 0.12 (0.05) & 0.68 (0.08) & 0.29 (0.08) & 0.75 (0.07) & 0.18 (0.05) & 0.67 (0.05)\\
$25, 25, 10$ & 0.25 (0.06) & 0.68 (0.06) & 0.14 (0.04) & 0.63 (0.05) & 0.63 (0.06) & 1.16 (0.05) & 0.44 (0.06) & 0.88 (0.08)\\ 
$25, 50, 10$ & 0.25 (0.06) & 0.69 (0.07) & 0.15 (0.04) & 0.65 (0.06) & 0.65 (0.07) & 1.22 (0.07) & 0.45 (0.06) & 0.92 (0.06)\\ 
$25, 100, 10$ & 0.26 (0.06) & 0.68 (0.07) & 0.15 (0.02) & 0.61 (0.05) & 0.63 (0.06) & 1.21 (0.08) & 0.42 (0.05) & 0.92 (0.07)\\
$25, 250, 10$ & 0.22 (0.05) & 0.68 (0.05) & 0.16 (0.03) & 0.62 (0.05) & 0.63 (0.06) & 1.24 (0.07) & 0.42 (0.06) & 0.92 (0.08)\\ 
$50, 25, 5$ & 0.11 (0.03) & 0.69 (0.06) & 0.09 (0.03) & 0.68 (0.06) & 0.18 (0.04) & 0.67 (0.04) & 0.11 (0.03) & 0.63 (0.04)\\ 
$50, 50, 5$ & 0.11 (0.03) & 0.69 (0.05) & 0.08 (0.03) & 0.68 (0.04) & 0.16 (0.03) & 0.67 (0.04) & 0.12 (0.02) & 0.65 (0.04)\\ 
$50, 100, 5$ & 0.10 (0.03) & 0.71 (0.07) & 0.08 (0.02) & 0.70 (0.08) & 0.16 (0.02) & 0.68 (0.05) & 0.12 (0.02) & 0.65 (0.04)\\ 
$50, 250, 5$ & 0.11 (0.03) & 0.71 (0.07) & 0.08 (0.02) & 0.69 (0.07) & 0.17 (0.03) & 0.68 (0.04) & 0.10 (0.02) & 0.64 (0.03)\\
$50, 25, 10$ & 0.15 (0.03) & 0.65 (0.05) & 0.11 (0.03) & 0.62 (0.04) & 0.44 (0.07) & 0.91 (0.06) & 0.26 (0.04) & 0.65 (0.05)\\ 
$50, 50, 10$ & 0.15 (0.04) & 0.65 (0.04) & 0.11 (0.03) & 0.62 (0.04) & 0.41 (0.05) & 0.94 (0.05) & 0.24 (0.04) & 0.66 (0.04)\\ 
$50, 100, 10$ & 0.14 (0.03) & 0.62 (0.04) & 0.10 (0.02) & 0.60 (0.04) & 0.39 (0.04) & 0.94 (0.05) & 0.22 (0.03) & 0.66 (0.03)\\ 
$50, 250, 10$ & 0.14 (0.03) & 0.62 (0.04) & 0.09 (0.02) & 0.59 (0.03) & 0.40 (0.05) & 0.96 (0.05) & 0.22 (0.04) & 0.67 (0.04)\\
$100, 25, 5$ & 0.08 (0.03) & 0.69 (0.06) & 0.05 (0.02) & 0.69 (0.06) & 0.12 (0.03) & 0.64 (0.04) & 0.08 (0.02) & 0.63 (0.04)\\ 
$100, 50, 5$ & 0.07 (0.03) & 0.71 (0.07) & 0.06 (0.01) & 0.71 (0.07) & 0.11 (0.02) & 0.64 (0.05) & 0.08 (0.03) & 0.63 (0.05)\\ 
$100, 100, 5$ & 0.07 (0.02) & 0.72 (0.04) & 0.05 (0.01) & 0.71 (0.04) & 0.09 (0.01) & 0.66 (0.03) & 0.07 (0.02) & 0.65 (0.03)\\ 
$100, 250, 5$ & 0.07 (0.03) & 0.70 (0.09) & 0.05 (0.01) & 0.69 (0.09) & 0.11 (0.02) & 0.64 (0.04) & 0.07 (0.01) & 0.63 (0.04)\\
$100, 25, 10$ & 0.11 (0.03) & 0.59 (0.04) & 0.08 (0.02) & 0.58 (0.04) & 0.25 (0.04) & 0.67 (0.04) & 0.16 (0.02) & 0.54 (0.03)\\ 
$100, 50, 10$ & 0.11 (0.03) & 0.61 (0.05) & 0.07 (0.01) & 0.60 (0.05) & 0.22 (0.03) & 0.66 (0.03) & 0.13 (0.01) & 0.52 (0.02)\\ 
$100, 100, 10$ & 0.10 (0.02) & 0.60 (0.04) & 0.07 (0.01) & 0.59 (0.04) & 0.22 (0.03) & 0.67 (0.04) & 0.14 (0.02) & 0.53 (0.02)\\ 
$100, 250, 10$ & 0.10 (0.02) & 0.61 (0.03) & 0.07 (0.02) & 0.60 (0.03) & 0.19 (0.03) & 0.66 (0.04) & 0.12 (0.02) & 0.54 (0.03)\\
$250, 25, 5$ & 0.05 (0.02) & 0.66 (0.07) & 0.04 (0.01) & 0.65 (0.07) & 0.07 (0.01) & 0.62 (0.03) & 0.05 (0.01) & 0.61 (0.03)\\ 
$250, 50, 5$ & 0.05 (0.01) & 0.69 (0.07) & 0.03 (0.01) & 0.68 (0.07) & 0.07 (0.02) & 0.62 (0.04) & 0.05 (0.01) & 0.61 (0.04)\\ 
$250, 100, 5$ & 0.04 (0.01) & 0.70 (0.10) & 0.03 (0.01) & 0.70 (0.10) & 0.06 (0.01) & 0.61 (0.04) & 0.05 (0.01) & 0.61 (0.05)\\ 
$250, 250, 5$ & 0.05 (0.01) & 0.68 (0.08) & 0.03 (0.01) & 0.67 (0.06) & 0.07 (0.01) & 0.62 (0.03) & 0.05 (0.01) & 0.61 (0.03)\\
$250, 25, 10$ & 0.07 (0.02) & 0.60 (0.04) & 0.05 (0.01) & 0.59 (0.05) & 0.13 (0.02) & 0.51 (0.02) & 0.08 (0.01) & 0.47 (0.02)\\ 
$250, 50, 10$ & 0.06 (0.02) & 0.58 (0.04) & 0.05 (0.01) & 0.58 (0.05) & 0.11 (0.02) & 0.50 (0.02) & 0.07 (0.01) & 0.46 (0.02)\\ 
$250, 100, 10$ & 0.06 (0.01) & 0.59 (0.04) & 0.04 (0.01) & 0.59 (0.04) & 0.10 (0.01) & 0.50 (0.02) & 0.07 (0.01) & 0.46 (0.01)\\ 
$250, 250, 10$ & 0.06 (0.02) & 0.59 (0.05) & 0.04 (0.01) & 0.59 (0.05) & 0.11 (0.02) & 0.50 (0.02) & 0.07 (0.01) & 0.46 (0.02)\\
       \midrule
        \bottomrule
     \end{tabular}
  \end{threeparttable}
  }
\end{table}

\noindent The results in Table \ref{tab:simressup1} show a pattern similar to the one observed in Table \ref{tab:simres2} of the main text: as $|\mathcal{M}_i|$ increases, parameter estimation (RMSE $\bm{b}$) and prediction (RMSE $\bm{R}$) improve, given that more observations become available. 

\begin{table}[H]
\resizebox{\textwidth}{!}{%
\centering
  \begin{threeparttable}
  \caption{Results for the proposed method on simulated data without latent variables with $k = 10$. The RMSE is averaged across 20 fitted models for each parameter combination, where the posterior mean was used as a point estimate. Standard deviations are provided between brackets.}
  \label{tab:simressup2}
     \begin{tabular}{c | cc | cc | cc | cc | cc | cc }
        \toprule
        \midrule
         & \multicolumn{6}{c|}{Linear} & \multicolumn{6}{c}{Bilinear}\\ \midrule
         & \multicolumn{2}{c|}{$|\mathcal{M}_i| = 1$}& \multicolumn{2}{c|}{$|\mathcal{M}_i| = 5$} & \multicolumn{2}{c|}{$|\mathcal{M}_i| = 10$} & \multicolumn{2}{c|}{$|\mathcal{M}_i| = 1$} & \multicolumn{2}{c|}{$|\mathcal{M}_i| = 5$} & \multicolumn{2}{c}{$|\mathcal{M}_i| = 10$}\\ \midrule
         \textbf{$n, m, p$} & \textbf{RMSE} $\bm{b}$ & \textbf{RMSE} $\bm{R}$ & \textbf{RMSE} $\bm{B}$ & \textbf{RMSE} $\bm{R}$ & \textbf{RMSE} $\bm{b}$ & \textbf{RMSE} $\bm{R}$ & \textbf{RMSE} $\bm{B}$ & \textbf{RMSE} $\bm{R}$ & \textbf{RMSE} $\bm{b}$ & \textbf{RMSE} $\bm{R}$ & \textbf{RMSE} $\bm{B}$ & \textbf{RMSE} $\bm{R}$\\ \midrule
$25, 25, 5$ & 0.31 (0.08) & 1.43 (0.15) & 0.11 (0.03) & 1.07 (0.11) & 0.08 (0.02) & 1.05 (0.11) & 0.66 (0.13) & 2.49 (0.38) & 0.20 (0.06) & 1.16 (0.11)  & 0.12 (0.02) & 1.03 (0.08) \\
$25, 50, 5$ & 0.27 (0.10) & 1.37 (0.22) & 0.11 (0.02) & 1.07 (0.09) & 0.09 (0.02) & 1.05 (0.08) & 0.57 (0.10) & 2.28 (0.28) & 0.19 (0.04) & 1.16 (0.09) & 0.11 (0.02) & 1.01 (0.06)\\ 
$25, 100, 5$ & 0.30 (0.12) & 1.45 (0.18) & 0.10 (0.02) & 1.09 (0.09) & 0.07 (0.02) & 1.05 (0.10) & 0.62 (0.11) & 2.41 (0.27) & 0.18 (0.04) & 1.14 (0.06) & 0.12 (0.02) & 1.04 (0.08)\\
$25, 250, 5$ & 0.30 (0.14) & 1.39 (0.18) & 0.11 (0.04) & 1.05 (0.11) & 0.08 (0.03) & 1.02 (0.12) & 0.64 (0.14) & 2.42 (0.35) & 0.20 (0.06) & 1.16 (0.10) & 0.12 (0.03) & 1.01 (0.07)\\
$25, 25, 10$ & 0.53 (0.11) & 2.09 (0.31) & 0.16 (0.05) & 1.02 (0.10) & 0.10 (0.03) & 0.96 (0.07) & 0.91 (0.07) & 3.64 (0.15) & 0.59 (0.07) & 2.42 (0.16) & 0.38 (0.06) & 1.61 (0.19)\\ 
$25, 50, 10$ & 0.46 (0.11) & 1.93 (0.29) & 0.14 (0.03) & 1.04 (0.09) & 0.10 (0.02) & 0.97 (0.09) & 0.92 (0.07) & 3.64 (0.14) & 0.61 (0.08) & 2.53 (0.19) & 0.37 (0.08) & 1.66 (0.18)\\ 
$25, 100, 10$ & 0.54 (0.14) & 2.11 (0.31) & 0.18 (0.05) & 1.02 (0.09) & 0.11 (0.02) & 0.92 (0.07) & 0.89 (0.07) & 3.68 (0.15) & 0.58 (0.06) & 2.50 (0.20) & 0.36 (0.06) & 1.70 (0.17)\\ 
$25, 250, 10$ & 0.48 (0.14) & 1.96 (0.28) & 0.15 (0.03) & 1.03 (0.06) & 0.11 (0.03) & 0.94 (0.08) & 0.91 (0.07) & 3.71 (0.11) & 0.59 (0.07) & 2.53 (0.21) & 0.35 (0.05) & 1.67 (0.18)\\
$50, 25, 5$ & 0.17 (0.05) & 1.16 (0.11) & 0.08 (0.03) & 1.04 (0.09) & 0.06 (0.02) & 1.02 (0.08) & 0.37 (0.10) & 1.69 (0.29) & 0.12 (0.03) & 1.02 (0.06) & 0.07 (0.02) & 0.96 (0.06)\\ 
$50, 50, 5$ & 0.20 (0.06) & 1.17 (0.11) & 0.08 (0.02) & 1.03 (0.07) & 0.05 (0.02) & 1.01 (0.06) & 0.35 (0.10) & 1.65 (0.25) & 0.11 (0.03) & 1.02 (0.06) & 0.08 (0.01) & 0.97 (0.05)\\ 
$50, 100, 5$ & 0.18 (0.05) & 1.22 (0.14) & 0.07 (0.02) & 1.07 (0.11) & 0.05 (0.02) & 1.06 (0.11) & 0.36 (0.08) & 1.66 (0.20) &  0.11 (0.02) & 1.02 (0.06) & 0.08 (0.02) & 0.97 (0.07)\\ 
$50, 250, 5$ & 0.18 (0.06) & 1.23 (0.17) & 0.07 (0.02) & 1.06 (0.11) & 0.05 (0.02) & 1.04 (0.11) & 0.35 (0.07) & 1.64 (0.17) & 0.11 (0.02) & 1.02 (0.07) & 0.07 (0.01) & 0.95 (0.05)\\
$50, 25, 10$ & 0.29 (0.10) & 1.41 (0.20) & 0.10 (0.02) & 0.98 (0.07) & 0.07 (0.02) & 0.93 (0.06) & 0.83 (0.07) & 3.36 (0.15) & 0.38 (0.08) & 1.70 (0.19) & 0.19 (0.03) & 1.06 (0.11)\\ 
$50, 50, 10$ & 0.25 (0.06) & 1.36 (0.13) & 0.10 (0.03) & 0.97 (0.05) & 0.07 (0.01) & 0.94 (0.06) & 0.82 (0.06) & 3.35 (0.14) & 0.34 (0.05) & 1.71 (0.13) & 0.17 (0.02) & 1.07 (0.08)\\ 
$50, 100, 10$ & 0.30 (0.10) & 1.47 (0.22) & 0.09 (0.03) & 0.94 (0.06) & 0.07 (0.01) & 0.91 (0.06) & 0.82 (0.05) & 3.44 (0.19) & 0.33 (0.05) & 1.75 (0.17) & 0.16 (0.02) & 1.07 (0.06)\\ 
$50, 250, 10$ & 0.31 (0.08) & 1.51 (0.21) & 0.09 (0.02) & 0.93 (0.06) & 0.07 (0.02) & 0.89 (0.05) & 0.83 (0.06) & 3.48 (0.10) & 0.33 (0.05) & 1.77 (0.13) & 0.15 (0.03) & 1.06 (0.07)\\
$100, 25, 5$ & 0.14 (0.08) & 1.12 (0.08) & 0.05 (0.01) & 1.04 (0.10) & 0.03 (0.01) & 1.03 (0.09) & 0.22 (0.05) & 1.21 (0.10) & 0.08 (0.02) & 0.97 (0.06) & 0.05 (0.01) & 0.94 (0.06)\\ 
$100, 50, 5$ & 0.11 (0.03) & 1.13 (0.10) & 0.05 (0.02) & 1.07 (0.10) & 0.04 (0.01) & 1.06 (0.10) & 0.21 (0.06) & 1.25 (0.13) &0.07 (0.01) & 0.97 (0.08) & 0.05 (0.01) & 0.94 (0.07)\\ 
$100, 100, 5$ & 0.11 (0.03) & 1.15 (0.08) & 0.04 (0.01) & 1.08 (0.06) & 0.03 (0.01) & 1.07 (0.06) & 0.18 (0.04) & 1.25 (0.06) & 0.06 (0.01) & 0.99 (0.05) & 0.05 (0.01) & 0.97 (0.05)\\ 
$100, 250, 5$ & 0.12 (0.03) & 1.13 (0.12) & 0.05 (0.02) & 1.05 (0.13) & 0.03 (0.01) & 1.04 (0.13) & 0.19 (0.05) & 1.24 (0.12) & 0.07 (0.01) & 0.96 (0.05) & 0.05 (0.01) & 0.94 (0.06)\\
$100, 25, 10$ & 0.19 (0.05) & 1.07 (0.12) & 0.07 (0.02) & 0.89 (0.07) & 0.05 (0.01) & 0.87 (0.07) & 0.67 (0.05) & 2.78 (0.10) & 0.19 (0.03) & 1.07 (0.05) & 0.10 (0.02) & 0.82 (0.05)\\ 
$100, 50, 10$ & 0.16 (0.04) & 1.09 (0.09) & 0.07 (0.02) & 0.91 (0.07) & 0.04 (0.01) & 0.90 (0.08) & 0.66 (0.04) & 2.88 (0.10) & 0.17 (0.03) & 1.05 (0.07) & 0.09 (0.01) & 0.80 (0.04)\\ 
$100, 100, 10$ & 0.15 (0.03) & 1.08 (0.07) & 0.07 (0.01) & 0.90 (0.06) & 0.04 (0.01) & 0.88 (0.06) & 0.64 (0.06) & 2.85 (0.11) & 0.16 (0.02) & 1.09 (0.09) & 0.09 (0.01) & 0.81 (0.03)\\ 
$100, 250, 10$ & 0.15 (0.04) & 1.11 (0.08) & 0.07 (0.02) & 0.92 (0.05) & 0.04 (0.01) & 0.90 (0.04) & 0.62 (0.05) & 2.94 (0.13) & 0.14 (0.02) & 1.08 (0.07) & 0.08 (0.01) & 0.82 (0.04)\\
$250, 25, 5$ & 0.08 (0.02) & 1.02 (0.11) & 0.03 (0.01) & 0.99 (0.11) & 0.02 (0.01) & 0.98 (0.11) & 0.12 (0.02) & 1.02 (0.04) & 0.05 (0.01) & 0.92 (0.04) & 0.03 (0.01) & 0.91 (0.05)\\ 
$250, 50, 5$ & 0.08 (0.02) & 1.06 (0.10) & 0.03 (0.01) & 1.03 (0.11) & 0.02 (0.01) & 1.03 (0.10) & 0.11 (0.03) & 1.01 (0.05) & 0.04 (0.01) & 0.93 (0.05) & 0.03 (0.01) & 0.92 (0.05)\\ 
$250, 100, 5$ & 0.07 (0.02) & 1.08 (0.14) & 0.03 (0.01) & 1.05 (0.14) & 0.02 (0.01) & 1.05 (0.14) & 0.11 (0.04) & 1.02 (0.09) & 0.04 (0.01) & 0.92 (0.07) & 0.03 (0.01) & 0.91 (0.07)\\ 
$250, 250, 5$ & 0.07 (0.01) & 1.03 (0.10) & 0.03 (0.01) & 1.00 (0.09) & 0.02 (0.01) & 1.00 (0.09) & 0.10 (0.02) & 1.01 (0.04) & 0.04 (0.01) & 0.92 (0.04) & 0.03 (0.01) & 0.91 (0.04)\\
$250, 25, 10$ & 0.10 (0.04) & 0.95 (0.06) & 0.04 (0.01) & 0.89 (0.06) & 0.03 (0.01) & 0.88 (0.07) & 0.37 (0.07) & 1.75 (0.15) & 0.09 (0.01) & 0.77 (0.03) & 0.06 (0.01) & 0.71 (0.03)\\ 
$250, 50, 10$ & 0.09 (0.02) & 0.93 (0.07) & 0.04 (0.01) & 0.87 (0.07) & 0.03 (0.01) & 0.86 (0.07) & 0.32 (0.05) & 1.71 (0.12) & 0.08 (0.01) & 0.76 (0.03) & 0.05 (0.01) & 0.69 (0.03)\\ 
$250, 100, 10$ & 0.08 (0.01) & 0.94 (0.07) & 0.03 (0.01) & 0.89 (0.06) & 0.03 (0.01) & 0.88 (0.06) & 0.30 (0.04) & 1.73 (0.16) & 0.07 (0.01) & 0.76 (0.03)  & 0.05 (0.01) & 0.70 (0.02)\\ 
$250, 250, 10$ & 0.09 (0.02) & 0.97 (0.07) & 0.04 (0.01) & 0.88 (0.07) & 0.03 (0.01) & 0.88 (0.07) & 0.30 (0.04) & 1.75 (0.17) & 0.07 (0.01) & 0.76 (0.04) & 0.04 (0.01) & 0.69 (0.03)\\
       \midrule
        \bottomrule
     \end{tabular}
  \end{threeparttable}
  }
\end{table}

\noindent Compared to the results shown in Table \ref{tab:simres1}, increasing $k$ results in a larger predictive error, as evidenced by Table \ref{tab:simressup1}. This is not surprising, given that the predictions can take on more values compared to the simulations of Table \ref{tab:simres1} with a smaller $k$.

\begin{table}[H]
\resizebox{\textwidth}{!}{%
\centering
  \begin{threeparttable}
  \caption{Results for the proposed method on simulated data with latent variables with $l = 1$, $k = 10$ and where the functional form of the predictor is linear. The RMSE is averaged across 20 fitted models for each parameter combination, where the posterior mean was used as a point estimate. Standard deviations are provided between brackets.}
  \label{tab:simressup3}
     \begin{tabular}{c | ccc | ccc | ccc }
        \toprule
        \midrule
         & \multicolumn{3}{c|}{$|\mathcal{M}_i| = 1$} & \multicolumn{3}{c}{$|\mathcal{M}_i| = 5$} & \multicolumn{3}{c}{$|\mathcal{M}_i| = 10$}\\ \midrule
         \textbf{$n, m, p$} & \textbf{RMSE} $\bm{b}$ & \textbf{RMSE} $\bm{F}$ & \textbf{RMSE} $\bm{R}$  & \textbf{RMSE} $\bm{b}$ & \textbf{RMSE} $\bm{F}$ & \textbf{RMSE} $\bm{R}$ & \textbf{RMSE} $\bm{b}$ & \textbf{RMSE} $\bm{F}$ & \textbf{RMSE} $\bm{R}$\\ \midrule
$25, 25, 5$ & 0.32 (0.11) & 2.80 (4.14) & 1.49 (0.17) & 0.11 (0.03) & 0.39 (0.27) & 1.11 (0.14) & 0.08 (0.03) & 0.27 (0.11) & 1.09 (0.13)\\ 
$25, 50, 5$ & 0.27 (0.09) & 3.86 (5.94) & 1.43 (0.18) & 0.12 (0.03) & 0.32 (0.12) & 1.12 (0.09) & 0.07 (0.02) & 0.33 (0.27) & 1.09 (0.09)\\ 
$25, 100, 5$ & 0.27 (0.10) & 11.01 (17.68) & 1.43 (0.12) & 0.11 (0.04) & 0.87 (0.94) & 1.13 (0.09) & 0.08 (0.02) & 0.67 (0.57) & 1.11 (0.09)\\ 
$25, 250, 5$ & 0.34 (0.11) & 6.95 (6.66) & 1.52 (0.17) & 0.12 (0.05) & 2.13 (2.78) & 1.11 (0.12) & 0.08 (0.02) & 4.76 (17.36) & 1.08 (0.12)\\ 
$25, 25, 10$ & 0.52 (0.11) & 2.40 (2.16) & 2.11 (0.26) & 0.16 (0.04) & 0.37 (0.25) & 1.07 (0.10) & 0.12 (0.03) & 0.29 (0.15) & 1.00 (0.11)\\ 
$25, 50, 10$ & 0.52 (0.11) & 11.42 (9.81) & 2.18 (0.24) & 0.15 (0.04) & 3.72 (15.03) & 1.13 (0.10) & 0.11 (0.02) & 0.36 (0.27) & 1.03 (0.10)\\ 
$25, 100, 10$ & 0.57 (0.16) & 14.32 (23.43) & 2.20 (0.30) & 0.18 (0.04) & 1.48 (2.67) & 1.06 (0.08) & 0.10 (0.02) & 0.37 (0.26) & 0.95 (0.08)\\ 
$25, 250, 10$ & 0.56 (0.15) & 74.46 (135.25) & 2.24 (0.28) & 0.17 (0.04) & 1.83 (2.84) & 1.09 (0.10) & 0.11 (0.02) & 0.67 (0.55) & 0.98 (0.08)\\ 
$50, 25, 5$ & 0.20 (0.07) & 2.38 (4.32) & 1.25 (0.10) & 0.08 (0.03) & 0.34 (0.20) & 1.11 (0.10) & 0.05 (0.01) & 0.28 (0.09) & 1.08 (0.10)\\ 
$50, 50, 5$ & 0.17 (0.03) & 9.69 (27.48) & 1.20 (0.08) & 0.08 (0.02) & 0.35 (0.29) & 1.07 (0.08) & 0.06 (0.02) & 0.36 (0.39) & 1.05 (0.07)\\ 
$50, 100, 5$ & 0.19 (0.05) & 5.66 (8.19) & 1.25 (0.12) & 0.07 (0.04) & 0.31 (0.15) & 1.11 (0.11) & 0.05 (0.02) & 0.52 (0.78) & 1.09 (0.11)\\ 
$50, 250, 5$ & 0.19 (0.05) & 6.70 (5.04) & 1.22 (0.10) & 0.08 (0.02) & 1.11 (1.27) & 1.09 (0.06) & 0.06 (0.01) & 0.30 (0.09) & 1.07 (0.07)\\ 
$50, 25, 10$ & 0.30 (0.09) & 4.26 (10.81) & 1.46 (0.17) & 0.11 (0.03) & 0.30 (0.17) & 1.01 (0.08) & 0.07 (0.02) & 0.24 (0.10) & 0.98 (0.08)\\ 
$50, 50, 10$ & 0.24 (0.05) & 10.80 (32.29) & 1.36 (0.13) & 0.09 (0.02) & 0.38 (0.49) & 1.00 (0.08) & 0.07 (0.01) & 0.31 (0.19) & 0.96 (0.06)\\ 
$50, 100, 10$ & 0.27 (0.06) & 7.21 (22.74) & 1.43 (0.20) & 0.09 (0.02) & 0.48 (0.41) & 0.98 (0.07) & 0.06 (0.02) & 0.29 (0.09) & 0.94 (0.06)\\ 
$50, 250, 10$ & 0.26 (0.05) & 11.10 (18.82) & 1.48 (0.14) & 0.09 (0.02) & 0.60 (0.41) & 0.98 (0.03) & 0.06 (0.01) & 0.66 (0.83) & 0.94 (0.04)\\ 
$100, 25, 5$ & 0.11 (0.03) & 2.05 (4.47) & 1.15 (0.10) & 0.05 (0.01) & 0.37 (0.23) & 1.08 (0.09) & 0.04 (0.01) & 0.29 (0.11) & 1.07 (0.09)\\ 
$100, 50, 5$ & 0.12 (0.03) & 2.04 (3.04) & 1.19 (0.11) & 0.05 (0.01) & 0.30 (0.08) & 1.11 (0.11) & 0.03 (0.01) & 0.38 (0.33) & 1.10 (0.11)\\ 
$100, 100, 5$ & 0.11 (0.03) & 3.55 (6.71) & 1.18 (0.08) & 0.05 (0.01) & 0.37 (0.15) & 1.12 (0.06) & 0.04 (0.01) & 0.29 (0.14) & 1.11 (0.07)\\ 
$100, 250, 5$ & 0.13 (0.02) & 6.48 (8.32) & 1.14 (0.10) & 0.05 (0.02) & 0.44 (0.33) & 1.05 (0.09) & 0.03 (0.01) & 0.37 (0.14) & 1.04 (0.09)\\ 
$100, 25, 10$ & 0.22 (0.06) & 0.91 (0.73) & 1.14 (0.09) & 0.08 (0.02) & 0.25 (0.08) & 0.92 (0.08) & 0.05 (0.01) & 0.24 (0.08) & 0.89 (0.08)\\ 
$100, 50, 10$ & 0.17 (0.04) & 4.07 (14.05) & 1.15 (0.08) & 0.07 (0.02) & 0.30 (0.13) & 0.95 (0.07) & 0.05 (0.01) & 0.25 (0.06) & 0.94 (0.07)\\ 
$100, 100, 10$ & 0.18 (0.05) & 11.17 (26.78) & 1.17 (0.13) & 0.07 (0.01) & 0.37 (0.15) & 0.94 (0.05) & 0.04 (0.01) & 0.45 (0.38) & 0.92 (0.06)\\ 
$100, 250, 10$ & 0.17 (0.03) & 4.88 (4.81) & 1.22 (0.16) & 0.05 (0.01) & 0.64 (0.46) & 0.96 (0.04) & 0.04 (0.01) & 0.34 (0.10) & 0.95 (0.05)\\ 
$250, 25, 5$ & 0.08 (0.03) & 0.53 (0.58) & 1.05 (0.11) & 0.04 (0.01) & 0.26 (0.09) & 1.02 (0.11)  & 0.02 (0.01) & 0.23 (0.05) & 1.02 (0.11)\\ 
$250, 50, 5$ & 0.08 (0.03) & 0.95 (1.12) & 1.10 (0.10) & 0.03 (0.01) & 0.37 (0.22) & 1.07 (0.11) & 0.03 (0.01) & 0.41 (0.39) & 1.07 (0.11)\\ 
$250, 100, 5$ & 0.08 (0.03) & 0.73 (0.99) & 1.11 (0.10) & 0.03 (0.01) & 0.29 (0.05) & 1.04 (0.08) & 0.02 (0.01) & 0.35 (0.14) & 1.03 (0.08)\\ 
$250, 250, 5$ & 0.07 (0.01) & 0.76 (0.72) & 1.05 (0.05) & 0.04 (0.01) & 3.72 (7.50) & 1.02 (0.05) & 0.02 (0.00) & 0.37 (0.21) & 1.02 (0.05)\\ 
$250, 25, 10$ & 0.11 (0.03) & 0.80 (1.47) & 1.01 (0.07) & 0.04 (0.01) & 0.34 (0.20) & 0.93 (0.08) & 0.03 (0.01) & 0.58 (1.25) & 0.92 (0.08)\\ 
$250, 50, 10$ & 0.10 (0.02) & 0.69 (0.71) & 0.97 (0.07) & 0.04 (0.01) & 0.37 (0.34) & 0.90 (0.08) & 0.03 (0.01) & 0.28 (0.12) & 0.90 (0.08)\\ 
$250, 100, 10$ & 0.09 (0.01) & 0.70 (0.47) & 0.98 (0.05) & 0.03 (0.01) & 2.89 (6.51) & 0.91 (0.05) & 0.03 (0.01) & 0.42 (0.36) & 0.90 (0.04)\\ 
$250, 250, 10$ & 0.11 (0.02) & 4.08 (6.63) & 0.99 (0.03) & 0.05 (0.01) & 0.40 (0.20) & 0.91 (0.04) & 0.03 (0.01) & 2.59 (5.49) & 0.90 (0.05)\\ 
       \midrule
        \bottomrule
     \end{tabular}
  \end{threeparttable}}
  %}
\end{table}

\noindent Similar to the model without latent factors, see Tables \ref{tab:simres1} and \ref{tab:simressup1}, increasing the value of $k$ also results in a larger predictive error for the model with latent factors, as shown in Table \ref{tab:simressup3}.

\begin{table}[H]
\resizebox{\textwidth}{!}{%
\centering
  \begin{threeparttable}
  \caption{Results for the proposed method on simulated data with latent variables with $l = 2$, $k = 5$ and where the functional form of the predictor is linear. The RMSE is averaged across 20 fitted models for each parameter combination, where the posterior mean was used as a point estimate. Standard deviations are provided between brackets.}
  \label{tab:simressup4}
     \begin{tabular}{c | ccc | ccc | ccc }
        \toprule
        \midrule
         & \multicolumn{3}{c|}{$|\mathcal{M}_i| = 1$} & \multicolumn{3}{c}{$|\mathcal{M}_i| = 5$} & \multicolumn{3}{c}{$|\mathcal{M}_i| = 10$}\\ \midrule
         \textbf{$n, m, p$} & \textbf{RMSE} $\bm{b}$ & \textbf{RMSE} $\bm{F}$ & \textbf{RMSE} $\bm{R}$  & \textbf{RMSE} $\bm{b}$ & \textbf{RMSE} $\bm{F}$ & \textbf{RMSE} $\bm{R}$ & \textbf{RMSE} $\bm{b}$ & \textbf{RMSE} $\bm{F}$ & \textbf{RMSE} $\bm{R}$\\ \midrule
$25, 25, 5$ & 0.42 (0.12) & 6.72 (11.39) & 0.91 (0.11) & 0.18 (0.05) & 0.80 (0.53) & 0.75 (0.07) & 0.12 (0.04) & 0.49 (0.30) & 0.71 (0.08)\\ 
$25, 50, 5$ & 0.41 (0.13) & 30.79 (66.97) & 0.95 (0.10) & 0.17 (0.05) & 1.08 (1.97) & 0.74 (0.05) & 0.12 (0.03) & 0.66 (0.65) & 0.73 (0.05)\\ 
$25, 100, 5$ & 0.41 (0.11) & 31.28 (43.11) & 0.96 (0.10) & 0.18 (0.05) & 3.88 (6.82) & 0.75 (0.07) & 0.12 (0.03) & 0.62 (0.49) & 0.73 (0.07)\\ 
$25, 250, 5$ & 0.45 (0.15) & 148.49 (549.85) & 0.94 (0.11) & 0.18 (0.06) & 4.48 (8.93) & 0.74 (0.08) & 0.12 (0.04) & 4.22 (8.16) & 0.71 (0.08)\\ 
$25, 25, 10$ & 0.59 (0.16) & 46.17 (132.28) & 1.10 (0.15) & 0.25 (0.07) & 0.66 (0.68) & 0.70 (0.05) & 0.17 (0.04) & 0.40 (0.19) & 0.64 (0.07)\\ 
$25, 50, 10$ & 0.57 (0.15) & 23.23 (28.54) & 1.15 (0.13) & 0.23 (0.05) & 0.73 (0.56) & 0.75 (0.07) & 0.15 (0.03) & 0.64 (0.64) & 0.67 (0.05)\\ 
$25, 100, 10$ & 0.67 (0.16) & 114.66 (369.45) & 1.20 (0.11) & 0.24 (0.06) & 1.90 (1.98) & 0.69 (0.05) & 0.17 (0.04) & 0.89 (1.05) & 0.63 (0.04)\\ 
$25, 250, 10$ & 0.62 (0.18) & 20.49 (25.41) & 1.15 (0.13) & 0.24 (0.07) & 4.90 (9.76) & 0.71 (0.06) & 0.14 (0.03) & 2.78 (4.57) & 0.64 (0.05)\\ 
$50, 25, 5$ & 0.28 (0.08) & 1.41 (1.45) & 0.80 (0.08) & 0.12 (0.04) & 0.49 (0.24) & 0.71 (0.07) & 0.09 (0.02) & 0.38 (0.12) & 0.71 (0.06)\\ 
$50, 50, 5$ & 0.27 (0.07) & 19.81 (45.6) & 0.81 (0.09) & 0.11 (0.03) & 0.57 (0.44) & 0.71 (0.05) & 0.08 (0.03) & 0.54 (0.30) & 0.70 (0.04)\\ 
$50, 100, 5$ & 0.28 (0.09) & 10.48 (18.84) & 0.83 (0.09) & 0.11 (0.03) & 1.80 (2.82) & 0.74 (0.07) & 0.07 (0.02) & 0.58 (0.48) & 0.72 (0.08)\\ 
$50, 250, 5$ & 0.23 (0.06) & 13.39 (12.41) & 0.78 (0.05) & 0.11 (0.03) & 1.52 (1.76) & 0.72 (0.05) & 0.08 (0.01) & 0.74 (0.37) & 0.71 (0.05)\\ 
$50, 25, 10$ & 0.38 (0.07) & 6.06 (7.43) & 0.89 (0.07) & 0.14 (0.04) & 0.82 (0.98) & 0.66 (0.06) & 0.11 (0.02) & 0.40 (0.13) & 0.64 (0.06)\\ 
$50, 50, 10$ & 0.38 (0.08) & 9.41 (18.12) & 0.93 (0.06) & 0.14 (0.02) & 0.66 (0.46) & 0.66 (0.04) & 0.10 (0.04) & 0.48 (0.19) & 0.63 (0.04)\\ 
$50, 100, 10$ & 0.43 (0.10) & 47.34 (74.36) & 0.95 (0.11) & 0.14 (0.02) & 1.25 (1.13) & 0.65 (0.04) & 0.10 (0.02) & 0.43 (0.13) & 0.62 (0.04)\\ 
$50, 250, 10$ & 0.43 (0.06) & 80.30 (107.27) & 0.97 (0.07) & 0.17 (0.02) & 1.99 (2.20) & 0.66 (0.02)  & 0.09 (0.02) & 0.61 (0.28) & 0.63 (0.02)\\ 
$100, 25, 5$ & 0.18 (0.04) & 13.86 (6.88) & 0.81 (0.06) & 0.08 (0.03) & 0.81 (0.80) & 0.74 (0.06) & 0.06 (0.01) & 0.52 (0.20) & 0.76 (0.04)\\ 
$100, 50, 5$ & 0.18 (0.03) & 10.47 (9.45) & 0.78 (0.06) & 0.07 (0.02) & 0.85 (0.92) & 0.73 (0.06) & 0.05 (0.01) & 0.61 (0.25) & 0.77 (0.05)\\ 
$100, 100, 5$ & 0.17 (0.04) & 18.19 (59.85) & 0.79 (0.05) & 0.07 (0.02) & 0.79 (0.86) & 0.74 (0.04) & 0.05 (0.01) & 0.47 (0.18) & 0.74 (0.04)\\ 
$100, 250, 5$ & 0.19 (0.07) & 16.95 (24.11) & 0.73 (0.05) & 0.08 (0.02) & 0.57 (0.26) & 0.69 (0.06) & 0.05 (0.01) & 0.62 (0.44) & 0.69 (0.06)\\ 
$100, 25, 10$ & 0.26 (0.06) & 16.00 (40.62) & 0.74 (0.05) & 0.10 (0.02) & 0.55 (0.21) & 0.63 (0.05) & 0.08 (0.02) & 0.84 (1.17) & 0.65 (0.04)\\ 
$100, 50, 10$ & 0.25 (0.06) & 10.67 (20.93) & 0.76 (0.06) & 0.09 (0.01) & 0.60 (0.24) & 0.63 (0.04) & 0.06 (0.01) & 0.89 (1.20) & 0.64 (0.04)\\ 
$100, 100, 10$ & 0.24 (0.06) & 18.19 (43.06) & 0.75 (0.04) & 0.09 (0.01) & 0.52 (0.19) & 0.62 (0.04) & 0.06 (0.01) & 0.77 (1.13) & 0.61 (0.04)\\ 
$100, 250, 10$ & 0.24 (0.06) & 25.61 (42.20) & 0.78 (0.06) & 0.09 (0.01) & 0.68 (0.20) & 0.64 (0.02) & 0.06 (0.01) & 0.45 (0.10) & 0.63 (0.03)\\ 
$250, 25, 5$ & 0.12 (0.05) & 1.02 (0.77) & 0.70 (0.08) & 0.05 (0.01) & 0.55 (0.26) & 0.68 (0.08) & 0.04 (0.02) & 0.63 (0.68) & 0.68 (0.08)\\ 
$250, 50, 5$ & 0.10 (0.03) & 1.14 (1.08) & 0.72 (0.08) & 0.05 (0.01) & 0.65 (0.56) & 0.71 (0.07) & 0.03 (0.01) & 0.66 (0.65) & 0.71 (0.07)\\ 
$250, 100, 5$ & 0.12 (0.03) & 1.23 (0.89) & 0.70 (0.05) & 0.06 (0.01) & 0.51 (0.19) & 0.69 (0.05) & 0.03 (0.01) & 0.41 (0.06) & 0.68 (0.05)\\ 
$250, 250, 5$ & 0.10 (0.04) & 1.13 (0.75) & 0.70 (0.05) & 0.05 (0.01) & 0.45 (0.19) & 0.68 (0.05) & 0.03 (0.00) & 0.61 (0.40) & 0.68 (0.03)\\ 
$250, 25, 10$ & 0.16 (0.04) & 1.44 (2.04) & 0.66 (0.04) & 0.07 (0.02) & 0.60 (0.68) & 0.62 (0.05) & 0.04 (0.01) & 0.49 (0.33) & 0.61 (0.04)\\ 
$250, 50, 10$ & 0.15 (0.03) & 2.73 (5.71) & 0.65 (0.05) & 0.06 (0.01) & 0.55 (0.33) & 0.60 (0.05) & 0.05 (0.01) & 0.53 (0.37) & 0.60 (0.05)\\ 
$250, 100, 10$ & 0.15 (0.03) & 1.90 (1.71) & 0.67 (0.04) & 0.06 (0.01) & 0.53 (0.19) & 0.61 (0.03) & 0.04 (0.01) & 0.37 (0.08) & 0.60 (0.04)\\ 
$250, 250, 10$ & 0.14 (0.03) & 1.64 (1.49) & 0.65 (0.04) & 0.06 (0.01) & 0.46 (0.15) & 0.60 (0.04) & 0.04 (0.00) & 0.55 (0.21) & 0.60 (0.03)\\ 
       \midrule
        \bottomrule
     \end{tabular}
  \end{threeparttable}}
  %}
\end{table}

\noindent Increasing $\ell = 1$ (Table \ref{tab:simres2}) to $\ell = 2$ (Table \ref{tab:simressup4}) does not result in worse parameter estimates for $\bm{b}$, but results in worse parameter estimates for $\bm{F}$ and slightly worse predictions. This makes sense, as the simulations for $\ell = 2$ require $nm$ more parameters to be estimated compared to those of $\ell = 1$.

\begin{table}[H]
\resizebox{\textwidth}{!}{%
\centering
  \begin{threeparttable}
  \caption{Results for the proposed method on simulated data with latent variables with $l = 2$, $k = 10$ and where the functional form of the predictor is linear. The RMSE is averaged across 20 fitted models for each parameter combination, where the posterior mean was used as a point estimate. Standard deviations are provided between brackets.}
  \label{tab:simressup5}
     \begin{tabular}{c | ccc | ccc | ccc }
        \toprule
        \midrule
         & \multicolumn{3}{c|}{$|\mathcal{M}_i| = 1$} & \multicolumn{3}{c}{$|\mathcal{M}_i| = 5$} & \multicolumn{3}{c}{$|\mathcal{M}_i| = 10$}\\ \midrule
         \textbf{$n, m, p$} & \textbf{RMSE} $\bm{b}$ & \textbf{RMSE} $\bm{F}$ & \textbf{RMSE} $\bm{R}$  & \textbf{RMSE} $\bm{b}$ & \textbf{RMSE} $\bm{F}$ & \textbf{RMSE} $\bm{R}$ & \textbf{RMSE} $\bm{b}$ & \textbf{RMSE} $\bm{F}$ & \textbf{RMSE} $\bm{R}$\\ \midrule
$25, 25, 5$ & 0.32 (0.09) & 2.83 (5.91) & 1.52 (0.25) & 0.12 (0.05) & 0.47 (0.15) & 1.16 (0.14) & 0.08 (0.02) & 0.36 (0.12) & 1.12 (0.15)\\ 
$25, 50, 5$ & 0.31 (0.10) & 13.92 (43.87) & 1.56 (0.16) & 0.12 (0.02) & 0.61 (0.43) & 1.17 (0.09) & 0.08 (0.02) & 0.50 (0.27) & 1.14 (0.09)\\ 
$25, 100, 5$ & 0.31 (0.10) & 8.65 (22.19) & 1.51 (0.21) & 0.12 (0.04) & 3.20 (7.49) & 1.18 (0.11) & 0.09 (0.03) & 0.56 (0.43) & 1.14 (0.11)\\ 
$25, 250, 5$ & 0.34 (0.15) & 91.33 (339.28) & 1.56 (0.25) & 0.12 (0.03) & 1.41 (1.46) & 1.14 (0.13) & 0.09 (0.04) & 4.44 (8.79) & 1.11 (0.12)\\ 
$25, 25, 10$ & 0.52 (0.11) & 11.13 (17.85) & 2.04 (0.21) & 0.18 (0.04) & 0.35 (0.12) & 1.10 (0.11) & 0.12 (0.03) & 0.47 (0.65) & 0.99 (0.12)\\ 
$25, 50, 10$ & 0.50 (0.15) & 27.62 (52.40) & 2.13 (0.34) & 0.16 (0.04) & 0.68 (0.72) & 1.17 (0.09) & 0.11 (0.03) & 0.58 (0.50) & 1.06 (0.11)\\ 
$25, 100, 10$ & 0.58 (0.16) & 102.87 (304.73) & 2.25 (0.25) & 0.17 (0.04) & 1.08 (1.90) & 1.10 (0.08) & 0.11 (0.03) & 0.65 (0.92) & 0.99 (0.08)\\ 
$25, 250, 10$ & 0.55 (0.16) & 31.83 (69.32) & 2.18 (0.34) & 0.17 (0.05) & 5.82 (9.71) & 1.10 (0.11) & 0.10 (0.02) & 1.19 (1.26) & 1.00 (0.08)\\ 
$50, 25, 5$ & 0.20 (0.06) & 2.49 (4.81) & 1.26 (0.13) & 0.09 (0.04) & 0.48 (0.32) & 1.12 (0.10) & 0.06 (0.01) & 0.36 (0.11) & 1.11 (0.09)\\ 
$50, 50, 5$ & 0.20 (0.06) & 13.29 (33.04) & 1.29 (0.15) & 0.08 (0.02) & 0.68 (0.73) & 1.11 (0.08) & 0.05 (0.01) & 0.60 (0.76) & 1.10 (0.08)\\ 
$50, 100, 5$ & 0.21 (0.07) & 5.15 (9.84) & 1.31 (0.16) & 0.07 (0.02) & 1.19 (2.12) & 1.15 (0.11) & 0.05 (0.01) & 0.48 (0.31) & 1.13 (0.12)\\ 
$50, 250, 5$ & 0.15 (0.04) & 5.83 (4.17) & 1.22 (0.06) & 0.08 (0.02) & 2.15 (2.78) & 1.14 (0.06) & 0.05 (0.01) & 0.64 (0.33) & 1.12 (0.07)\\ 
$50, 25, 10$ & 0.31 (0.06) & 4.18 (10.45) & 1.46 (0.14) & 0.11 (0.03) & 0.51 (0.66) & 1.04 (0.10) & 0.07 (0.02) & 0.37 (0.13) & 1.01 (0.12)\\ 
$50, 50, 10$ & 0.28 (0.05) & 1.71 (1.91) & 1.50 (0.11) & 0.10 (0.02) & 0.50 (0.32) & 1.03 (0.06) & 0.07 (0.02) & 0.52 (0.32) & 1.00 (0.07)\\ 
$50, 100, 10$ & 0.33 (0.08) & 49.00 (180.80) & 1.62 (0.22) & 0.10 (0.02) & 1.23 (2.17) & 1.02 (0.07) & 0.07 (0.02) & 0.73 (1.09) & 0.98 (0.06)\\ 
$50, 250, 10$ & 0.31 (0.09) & 20.24 (24.30) & 1.58 (0.26) & 0.12 (0.03) & 0.66 (0.39) & 1.03 (0.02) & 0.06 (0.02) & 0.41 (0.09) & 0.98 (0.02)\\ 
$100, 25, 5$ & 0.14 (0.03) & 3.40 (4.48) & 1.20 (0.08) & 0.06 (0.01) & 0.70 (0.86) & 1.12 (0.08) & 0.04 (0.02) & 0.33 (0.18) & 1.19 (0.08)\\ 
$100, 50, 5$ & 0.13 (0.04) & 24.75 (144.5) & 1.24 (0.07) & 0.07 (0.02) & 0.94 (1.00) & 1.14 (0.09) & 0.03 (0.01) & 0.76 (0.55) & 1.16 (0.06)\\ 
$100, 100, 5$ & 0.12 (0.03) & 33.53 (144.5) & 1.24 (0.07) & 0.05 (0.01) & 0.62 (0.72) & 1.16 (0.06) & 0.03 (0.01) & 0.53 (0.34) & 1.15 (0.06)\\ 
$100, 250, 5$ & 0.14 (0.03) & 7.11 (8.52) & 1.14 (0.10) & 0.06 (0.01) & 1.02 (0.84) & 1.08 (0.10) & 0.03 (0.01) & 0.47 (0.17) & 1.07 (0.10)\\ 
$100, 25, 10$ & 0.18 (0.05) & 2.12 (5.27) & 1.21 (0.13) & 0.07 (0.02) & 1.27 (2.03) & 1.02 (0.06) & 0.05 (0.02) & 0.60 (0.23) & 1.03 (0.06)\\ 
$100, 50, 10$ & 0.18 (0.05) & 5.33 (13.18) & 1.24 (0.16) & 0.06 (0.02) & 1.54 (2.68) & 1.05 (0.07) & 0.04 (0.01) & 0.51 (0.19) & 0.97 (0.06)\\ 
$100, 100, 10$ & 0.17 (0.04) & 15.01 (39.52) & 1.19 (0.09) & 0.06 (0.01) & 1.01 (1.49) & 0.97 (0.06) & 0.04 (0.01) & 0.43 (0.18) & 0.95 (0.06)\\ 
$100, 250, 10$ & 0.16 (0.04) & 12.57 (22.67) & 1.20 (0.10) & 0.06 (0.01) & 1.67 (2.34) & 1.00 (0.04) & 0.04 (0.01) & 0.43 (0.07) & 0.99 (0.05)\\ 
$250, 25, 5$ & 0.08 (0.03) & 0.84 (0.48) & 1.10 (0.13) & 0.04 (0.01) & 0.57 (0.60) & 1.07 (0.13) & 0.03 (0.01) & 0.54 (0.38) & 1.05 (0.12)\\ 
$250, 50, 5$ & 0.07 (0.02) & 0.98 (0.85) & 1.13 (0.11) & 0.03 (0.01) & 0.56 (0.40) & 1.11 (0.12) & 0.02 (0.01) & 0.44 (0.15) & 1.10 (0.11)\\ 
$250, 100, 5$ & 0.08 (0.03) & 9.17 (19.07) & 1.10 (0.07) & 0.04 (0.01) & 2.53 (4.82) & 1.07 (0.07) & 0.02 (0.01) & 0.85 (1.09) & 1.07 (0.06)\\ 
$250, 250, 5$ & 0.07 (0.02) & 0.97 (0.61) & 1.10 (0.10) & 0.04 (0.01) & 2.19 (2.93) & 1.07 (0.05) & 0.02 (0.00) & 0.67 (0.25) & 1.06 (0.05)\\ 
$250, 25, 10$ & 0.11 (0.03) & 1.27 (2.28) & 1.04 (0.08) & 0.05 (0.01) & 0.74 (0.85) & 0.97 (0.08) & 0.03 (0.01) & 0.39 (0.15) & 0.95 (0.07)\\ 
$250, 50, 10$ & 0.09 (0.02) & 3.80 (9.32) & 1.01 (0.07) & 0.04 (0.01) & 0.46 (0.18) & 0.94 (0.07) & 0.03 (0.01) & 0.52 (0.19) & 0.93 (0.08)\\ 
$250, 100, 10$ & 0.10 (0.02) & 2.58 (2.93) & 1.04 (0.07) & 0.04 (0.01) & 0.48 (0.14) & 0.95 (0.06) & 0.03 (0.01) & 0.38 (0.06) & 0.94 (0.06)\\ 
$250, 250, 10$ & 0.09 (0.02) & 0.98 (0.74) & 1.01 (0.07) & 0.04 (0.01) & 0.35 (0.11) & 0.94 (0.05) & 0.03 (0.00) & 0.50 (0.13) & 0.93 (0.04)\\ 
       \midrule
        \bottomrule
     \end{tabular}
  \end{threeparttable}}
  %}
\end{table}

\noindent As observed for all other simulations, increasing $k$ leads to larger predictive errors for $\ell = 2$, see Table \ref{tab:simressup5}. Additionally, similar to the case where $k = 5$, increasing $\ell$ from 1 to 2, also leads to worse estimates for $\bm{F}$ and worse predictions for $\bm{F}$ whenever $k = 10$.

\subsection*{Details of the real data simulations}
Table \ref{tab:simressup6} provides information on which covariates were used in the simulation study on the Amazon music, Goodbooks, MovieLens and Yelp datasets.

\begin{table}[H]
\resizebox{\textwidth}{!}{%
\centering
  \begin{threeparttable}
  \caption{Table containing the user and item covariates for all four datasets used in the simulations with real-world data.}
  \label{tab:simressup6}
     \begin{tabular}{c | c | c }
        \toprule
        \midrule
         \textbf{Data} & $\bm{X}$ \textbf{covariates} & $\bm{Y}$ \textbf{covariates}\\ \midrule
Amazon music & Average rating, number of rated items & Average rating, times rated, price, similar bought items\\ 
Goodbooks & Average rating, number of rated items, per genre ratios of items evaluated by a user that fit into that genre & Average rating, times rated, publication year, number of authors, genre\\ 
MovieLens & Average rating, number of rated items, occupation, gender, per genre ratios of items evaluated by a user that fit into that genre  & Average rating, times rated, release year, budget, running time, box office, genre\\ 
Yelp & Average rating, number of rated items, number of years elite, number of useful comments, number of funny comments, number of cool comments, number of fans & Average rating, times rated, price, atmosphere, type\\ 
       \midrule
        \bottomrule
     \end{tabular}
  \end{threeparttable}}
  %}
\end{table}

\end{document}